# Measuring User Perceived Security of Mobile Banking Applications

Richard Apau[a], Harjinder Singh Lallie[b*]

[a] *Lecturer, Department of Information Technology Education, Akenten Appiah- Menka University of Skills Training and Entreprenuerial Development (AAMUSTED), P.O. Box 1277, Kumasi Ghana*
[b] *Associate professor, WMG, University of Warwick, Coventry, CV4 7AL*
[*] corresponding author

**Abstract**

Mobile banking applications have gained popularity and have significantly revolutionised the banking industry. Despite the convenience offered by M-Banking Apps, users are often distrustful of the security of the applications due to an increasing trend of cyber security compromises, cyber-attacks, and data breaches. Considering the upsurge in cyber security vulnerabilities of M-Banking Apps and the paucity of research in this domain, this study was conducted to empirically measure user-perceived security of M-Banking Apps. A total of 315 responses from study participants were analysed using covariance-based structural equation modelling (CB-SEM). The results indicated that most constructs of the baseline Extended Unified Theory of Acceptance and Use of Technology (UTAUT2) structure were validated. Perceived security, institutional trust and technology trust were confirmed as factors that affect user's intention to adopt and use M-Banking Apps. However, perceived risk was not confirmed as a significant predictor. The current study further revealed that in the context of M-Banking Apps, the effects of security and trust are complex. The impact of perceived security and institutional trust on behavioural intention was moderated by age, gender, experience, income, and education, while perceived security on use behaviour was moderated by age, gender, and experience. The effect of technology trust on behavioural intention was moderated by age, education, and experience. Overall, the proposed conceptual model achieved acceptable fit and explained 79% of the variance in behavioural intention and 54.7% in use behaviour of M-Banking Apps, higher than that obtained in the original UTAUT2. The guarantee of enhanced security, advanced privacy mechanisms and trust should be considered paramount in future strategies aimed at promoting M-Banking Apps adoption and use.

## 1. Introduction

The growth in technological innovation, the increasing need for up-to-date information and the requirement for information systems that are convenient have significant importance in the banking industry and the financial sector as a whole (Majumdar & Pujari, 2021). Many banks, both traditional (e.g. Barclays, HSBC, Santander, NatWest etc.) and modern digital banks (e.g. Monzo, Monese, Starling etc.) have deployed mobile banking applications ("M-Banking Apps") to reduce operational and managerial cost while providing improvement in customer services and satisfaction.

The introduction of mobile banking has necessitated a rapid increase and popularity of mobile financial web apps (Singh and Srivastava, 2020; Shareef et al., 2018), which to a large extent offer many advantages to bank entities in terms of ease and comfort of performing transaction by clients, improving service quality as well as increasing market coverage (Rabaa'I & AlMaati, 2021; Namahoot & Laohavichien, 2018).

Mobile devices have become an integral part of our daily lives. Given the prevalence of mobile devices, M-Banking and online channels have become the preferred medium for users to access their banking services ( To & Trinh, 2021). As mobile devices become more versatile, they play a key role in the interaction between users and financial services providers.

Notwithstanding the prevalence of mobile devices in recent times, there are inherent risk and security vulnerabilities which could be exploited. The popularity of these devices has seen a corresponding increase in the appetite of cybercriminals and computer hackers to target such devices (Amro and Tiantian, 2017). Therefore, despite the convenience offered by M-Banking Apps, users could be distrustful of the security of the application (Ramli et al., 2021), and ultimately may want to stay with traditional banking services and methods.

Financial transactions through M-Banking Apps are characterised by uncertainty, anonymity and often without human contact or inter-personal relationship. This in itself creates circumstances of security threats to the user, thereby justifying their mistrust of M-Banking security within the virtual environment (Mettouris et al., 2015). For online users, security is a highly rated issue in their trust of mobile financial services (Kumar & Yukita, 2021). This is because users provide their banking details and other personal information while performing online financial transactions. The perception of security and risks of M-Banking Apps is a key aspect of the usage of the channel for mobile and online financial transactions. Many studies have been conducted to examine users perception of trust, risk and security of online banking, mobile banking and internet banking in general (Obaid, 2021; Kumar & Yukita, 2021; Merhi et al., 2021; Merhi et al., 2019; Mutahar *et al.*, 2018; Lim *et al.*, 2019). These studies have offered varied outcomes and analysis pertaining to the security threats of M-Banking.

Although issues of trust, security and risk of mobile banking have been extensively studied, particularly in developed countries, they failed to examine how individual differences and characteristics moderate the effects of perceived security, risk, and trust. While users of M-Banking Apps may become increasingly concerned about the safety and security of their transactions and data protection, the perception may differ among various demographics (Chawla and Joshi, 2018; Choudrie, Junior, McKenna, and Richter, 2018). For instance, Natarajan et al. (2018) have indicated that, intention to use electronic banking is more applicable to females than males and that older people harbour more trust towards electronic banking services than younger generations. There is a dearth of research examining the characteristics of age, gender, income, experience, and education level on the antecedents of security, risk, and trust. In many cases however, the moderating effects of demographic characteristics on the security perception of M-Banking Apps users are anecdotal rather than empirical. More importantly, different personality and geographic characteristics can impact security perceptions culminating into low adoption rate for M-Banking Apps.

Against this background, the key objective of this research is to measure how characteristics such as age, gender, education, income, and experience moderate the effect of perceived security on M-Banking Apps. Although the UTAUT and UTAUT2 as well as other studies (such as Bapista & Oliveira, 2015) examined how some characteristics moderate the effect of certain antecedents on adoption, how these characteristics moderate the effect of perceived security have not been investigated within the context of M-Banking Apps. Liébana-Cabanillas et al. (2014) found that user behavior differ across mobile payment platforms, hence it is possible that perceptions of security on M-Banking Apps may not replicate in other systems (Ramli et al., 2021; To & Trinh, 2021).

The present study is an effort to fill the gap in literature by addressing the shortcomings that have been identified. The outcome of this study will make significant theoretical and practical contributions. From a theoretical perspective, this study contributes to extant technology adoption literature as recommended by Venkatesh et al., (2012) through the expansion of extended unified theory of acceptance and use of technology (UTAUT2) model beyond the initial context for which it were conceived. As such, this study integrates four constructs, security, risk, institutional trust, and technology trust into the UTAUT2 with the aim of expanding the generalisability and the applicability of the theory in the M-Banking Apps context. This study is the first attempt at examining how demographic characteristics such as age, gender, income, education, and experience moderate perceived security in the context of M-Banking Apps. Again, while many of the existing study have examined trust in the broader perspective, this study departs from such studies and delineate trust into institutional trust and technology trust. Since trust is important determinant of online purchases and transactions, it is important to distinguish users trust in the internet medium (technology trust) and trust of banking institutions (institutional trust). The outcome of the study will facilitate the formulation of strategies aimed at ensuring the successful security implementation apps developers and also offer policy makers in the banking industry the opportunity to empirically examine consumers security concerns and put in place the required mitigating mechanisms.

The rest of this paper is structured as follows. Section Two presents the literature review comprising related works, theoretical framework, conceptual model for the study and hypothesis formulation, this is followed by the study methodology adopted in Section Three. Section Four present detailed discussion on the statistical analysis, results, and findings. The paper concludes with Section Five, giving elaboration on the study's main findings, conclusions, main contribution of the study as well as limitations and future research directions.

## 2. Literature Review

### 2.1. Related Works

Users perception of security is one of the main factors affecting many electronic applications including e-commerce, mobile banking, internet banking, online shopping among many other systems (Merhi et al., 2021). As human errors in online transactions are inevitable (Karjaluoto et al., 2021), several factors also come to play to compromise the mobile banking security as technological vulnerabilities (eg. cyber security threats) continue to be major concerns for M-Banking App deployment (Rehman & Shaikh, 2020; Prabhakaran et al., 2020). These concerns exist despite the advances in technology protection and security (Win et al., 2021; Tham *et al.*, 2017). Lack of information on M-Banking security has contributed to users' lack of awareness about security which has further heightened their perception of security and lack of trust in M-Banking services Obaid, 2021; Majumdar & Pujari, 2021). Notwithstanding the advances in internet and security technology, the fact that users do not have adequate information on M-Banking security measures have affected their appreciation of such systems (Mutahar *et al.*, 2018; Stewart & Jürjens, 2018). Security and trust are considered multidimensional factors that are constantly changing as banking transactions continue to rapidly shift from the physical world to the virtual world. Due to this, perceived security, perceived risk and trust have become the main antecedents of consumer acceptance and use of M-Banking Apps (Apau & Koranteng, 2019). Several studies have examined antecedents of security, trust, and risk of M-Banking.

In one of the earliest studies into the subject, Chen (2013) carried out an empirical investigation into perceived risk, usage frequency and brand awareness of mobile banking services. The study concluded that individuals' behavioural intention to use mobile banking services is influenced by risk perception. Svilar and Zupančič (2016) also found that perceived security is the most important factor in mobile and internet banking services. Muñoz-Leiva et al (2017) investigated the determinants to use mobile banking apps. The study suggested that trust affects user attitude and their perceived risk towards M-Banking Apps. However, perceived risk did not affect their intention to use mobile banking applications. Similarly, Tham *et al.*, (2017) and Pentina et al. (2016) revealed that users' perceived security and trust influence online banking adoption intention. Williams (2018) investigated payment and security perceptions of social commerce and mobile payment platforms. Williams' findings show that trust affects perceived risk. However, neither trust nor perceived risk had a significant effect on behavioural intention to use mobile payment platforms. In a related study, Stewart and Jürjens (2018) indicated that security affects trust and use intention of financial technologies.

Many other studies have empirically confirmed the effect of security (Rahi and Abd. Ghani, 2018; Lim *et al.,* 2019; Singh & Srivastava, 2020), risk perception (Namahoot and Laohavichien, 2018; Shaw & Sergueeva, 2019; Alonso-Dos-Santos *et al.,* 2020) and trust (Chauhan et al., 2021; Merhi *et al*., 2019; Alonso-Dos-Santos *et al.,* 2020; Widodo et al.. 2019) on the adoption and use of mobile banking services. More recently, Hanif and Lallie (2021) conducted an empirical study examining security factors on the intention to use mobile banking applications in the UK older generation (55+). Hanif and Lallie found that perceived cybersecurity risk influence intention to use mobile banking apps, but perceived cyber security trust and overall security had no effects on intention to use mobile banking apps among UK's older population. Chauhan et al (2021) also examined the effect of security and risk on the adoption of electronic banking services in India. The study results showed that security risk play important role in user intention to adopt and use electronic banking services.

In addition to security, risk, and trust, individual characteristics have been found to be important in mobile banking applications. While Arenas-Gaitán et al (2015) argued that gender plays no role in technology acceptance, other studies (Wang et al., 2016; Chawla and Joshi, 2018) have considered gender as an important variable in terms of mobile banking adoption. Also age, experience, income and educational have separately been identified in studies (Liébana-Cabanillas et al, 2014; Baptista & Oliveira, 2015; Laukkanen, 2015) as important influencers of technology adoption. In a recent study, Merhi et al (2021) improved upon an earlier research work by Merhi et al (2019) by integrating age and gender as a moderating characteristic of security on behavioural intention to use mobile banking apps. The study found that age moderates trust among UK consumers, but gender had no significant relationship with security, trust and privacy. User behavior differ across mobile payment platforms (Majumdar & Pujari, 2021; Widyanto et al., 2020; Liébana-Cabanillas et al., 2014), hence it is possible that perceptions of security on M-Banking Apps may not replicate in other systems. Although the UTAUT and UTAUT2 as well as other studies (e.g., Bapista & Oliveira, 2015) examined how some characteristics moderate the effect of certain antecedents on adoption, how these characteristics moderate the effect of perceived security, perceived trust (institution and technology) and risk have not been investigated within the context of M-Banking Apps.

**2.2 Theoretical Framework**

Several factors that influence user decision to adopt and use technology have been proposed in the academic literature. These have resulted in the proposition and development of theories on technology acceptance among organisations and individual consumers. These theories include the theory of reasoned action (TRA), theory of planned behaviour (TPB), technology acceptance model (TAM), and more recently, the extended unified theory of acceptance and use of technology (UTAUT2). The UTAUT2 is adopted as the theoretical foundation to form and guide the conduct of this study.

Venkatesh et al. (2012) proposed the extended unified theory of acceptance and use of technology (UTAUT2) as an extension of the unified theory of acceptance and use of technology (UTAUT) which was developed to predict organisations' acceptance of technology (Venkatesh, Morris, Davis, & Davis, 2003). The UTAUT distilled and brought to the fore, critical factors and contingencies that relate to the prediction of behavioural intention to accept and the use of technology focusing mainly in the context of organisation (Venkatesh et al., 2003). Since its original publication, the UTAUT has been used as a baseline model and has been applied in numerous studies evaluating varieties of technologies in both organisational and non-organisational settings. In addition to the popularity the model enjoyed, it has also received many applications which have contributed immensely to fortifying its generalisability (Venkatesh et al., 2012).

Over the years, technology acceptance has received a lot of recognition amongst researchers, as evidenced by the existence of numerous technology acceptance models and research works. M-Banking and its related applications have not been an exception as their acceptance has been predicted using five models (Merhi et al., 2019). These models include; the diffusion of innovation theory (DOI) (Rogers, 1995), theory of reasoned action (TRA) (Fishbein & Ajzen, 1975), theory of planned behaviour (TPB) (Ajzen, 1991), technology acceptance models (TAM) (Davis, 1989), and theory of perceived risk (TPR) (Featherman & Pavlou, 2003). These models were however, not without limitations. In order to overcome the limitation of the previously existing models, Venkatesh et al. (2003) proposed the UTAUT. The UTAUT model consisted of eight previously developed technology acceptance models. The UTAUT model was developed from four of the five models previously indicated and excluded the Theory of Perceived Risk (TPR). The four other models consolidated into the previous models to form the UTAUT include; the motivational model (MM) (Davis, Bagozzi, & Warshaw, 1992), the PC utilisation model (MPCU) (Thompson, Higgins & Howell, 1991), the social cognitive theory (SCT) (Bandura, 1986), as well as the integrated model of technology acceptance and planned behaviour (TAM-TPB) (Taylor & Todd, 1995).

The UTAUT model consisted of four constructs. These include *performance expectancy, effort expectancy*, *social influence* and *facilitating conditions*. The model accurately predicted system usage and acceptance in an organisational setting and also accounted for other influences as well as interaction terms such as demographic variables (age, gender, experience) (Venkatesh, Brown, & Bala, 2013). This made the UTAUT more efficacious in technology acceptance over its counterparts (Venkatesh and Zhang, 2010). UTAUT was applied in numerous scenarios across different countries and industries including for the prediction of internet technology (Servidio, 2014; Touray, 2015) internet banking (Martins, Oliveira and Popovič, 2014a), mobile banking (Afshan and Sharif, 2016; Bhatiasevi, 2016) and e-government (Rodrigues, Sarabdeen and Balasubramanian, 2016), among many other fields. Notwithstanding the success of UTAUT, it faced its own limitations as the conception was purely based on organisational settings rather than individual consumers (Negahban and Chung, 2014). Due to this limitation, there was the need to adapt and modify the model to better represent consumer acceptance and intention to use technology, especially e-business and e-commerce (Chen and Holsapple, 2013). In addition to the previously mentioned limitation, the UTAUT suffered other shortcomings such as task-technology fit, technology performance and user satisfaction which are pre-requisites for the measurement of technology usage and success (Montesdioca and Maçada, 2015).

Recognising the shortcomings of the UTAUT, Venkatesh et al. (2012) modified and extended the model by introducing three new constructs; *hedonic motivation*, *price value,* and *habit*, aimed at integrating the consumer context to address the limitations of the UTAUT (Merhi et al., 2019). The result of this was UTAUT2. The introduction of hedonic motivation as a new construct is considered a strong predictor that emphasises utility which was missing in the initial UTAUT. Unlike organisations, 'cost' is an important component of the consumers decision to use technology, hence the integration of price value. Although 'behavioural intention' has long been used to predict technology use, a new theorised construct called 'habit' is also considered another critical predictor for technology acceptance and use (Venkatesh et al, 2012), hence its integration as a construct. Thus, the introduction of habit highlights intention as an overarching mechanism and key driver of behaviour.

Beyond the addition of three constructs to the initial UTAUT, the UTAUT2 dropped voluntariness and rather proposed new relationships and moderators. For instance, a link was introduced between facilitating conditions and behavioural intentions. Although moderators were part of the UTAUT, the UTAUT2 proposed new moderators which were tested to be valid. The initial UTAUT had performance expectancy moderated by age and gender; effort expectance moderated by age, gender, and experience; social influence was moderated by age, gender, and experience; and facilitating condition to use behaviour was moderated by age and experience. UTAUT2 included a moderated relationship of age, gender, and experience pertaining to the three new constructs integrated.

The emergence of the UTAUT2 has facilitated and enhanced the concept of analysing consumer technology usage in a voluntary setting. UTAUT2 model showed higher superiority in the prediction of technology acceptance by achieving higher explanation of the variance in intention and use of technology (Venkatesh et al., 2012). Since its' introduction, UTAUT2 has received validation and extension in multiple fields including mobile banking (Baptista and Oliveira, 2015; Alalwan, Dwivedi and Rana, 2017; Merhi, Hone and Tarhini, 2019), internet banking (Alalwan, Dwivedi, Rana, Lal, & Williams, 2015), e-services (Fakhoury and Aubert, 2017), online shopping/ e-commerce (Fakhoury and Aubert, 2017; Tarhini, Alalwan, & Algharabat, 2019), e-learning (El-Masri & Tarhini, 2017), and usage of smartphone (Ahmed,

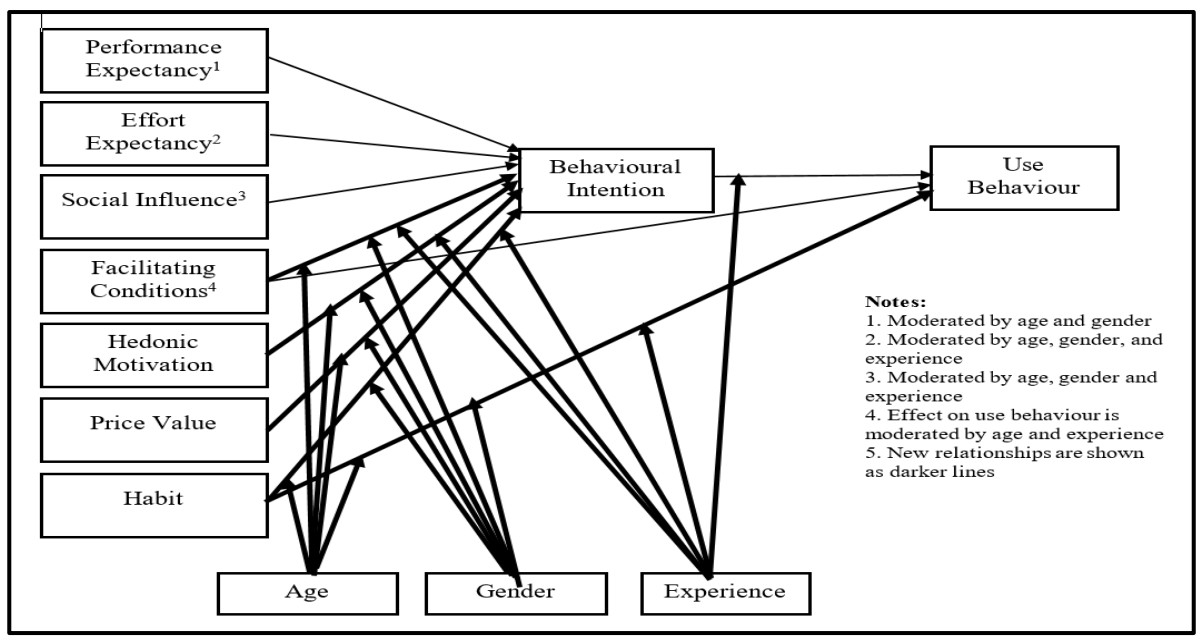

Figure 1 Original UTAUT2 Model (Venkatesh, Thong, & Xu, 2012)

Everatt and Fox-Turnbull, 2017; Ameen, Willis and Hussain Shah, 2018). These studies and validations have further established UTAUT2 as a more comprehensive framework for the evaluation technology acceptance and use in organisational and consumer context alike. The UTAUT2 is therefore adopted as the theoretical foundation for this study. **Figure 1** shows detail constructs and the relationship among the various constructs in UTAUT2.

### 2.3. Conceptual Model for the Study

The conceptual framework for this study uses the extended unified theory of acceptance and use of technology (UTAUT2) as its theoretical foundation and extends it with relevant constructs. Many studies have used TAM as the base model in investigating the factors facilitating internet or mobile banking adoption. Although TAM has extensively been used to predict user's technology adoption behaviour, the model has been overtaken by UTAUT2. The TAM theory excludes more important factors that explain an individual's technology acceptance such as their enjoyment, price value, habit, cultural influence, and demographic variables. The UTAUT2 model has shown higher superiority in the prediction of technology acceptance by achieving higher explanation of the variance in intention and use of technology.

While, the UTAUT2 has been validated in several industries and research settings, it falls short of constructs that have become relevant in the use of technology, particular, within the banking sector where risk is considered an important factor. The banking industry faces cyber security challenges (Wechuli, Franklin and Jotham, 2017). These challenges make factors such as security, risk, and trust important consideration by consumers to use M-Banking Apps. Consequently, this study extends the UTAUT2 with perceived security, perceived risk, and trust (institution and technology) as relevant constructs for the prediction of M-Banking Apps use by consumers. In addition to the above-mentioned constructs, the study will also examine how demographic characteristics (age, gender, educational level, income, and experience) act as moderators of user perceived security of M-Banking Apps. The present study will explore the relationship between the extended constructs and behavioural intentions on the one hand and between the extended constructs and user behaviour on the other hand with the demographic characteristics moderating the relationships. **Figure 2** demonstrates the proposed conceptual model.

### 2.4. Hypotheses Formulation

This research seeks to answer the question on how characteristics such as age, gender, education, income, and experience moderate the effect of perceived security, perceived risk, institutional trust, and technology trust on M-Banking Apps. The following hypotheses have been developed with the aim of answering the main research question in this study.

*Performance Expectancy (PE)*

Performance expectancy is defined as *"the degree to which an individual believes that using a system will help him or her to attain job performance"* (Venkatesh *et al.*, 2003). In effect, it is the benefit gained from using a technological innovation or system. This was considered as perceived usefulness in TAM (Davis, 1989), and relative advantage in DOI (Rogers, 1995). Performance expectancy suggests that individual will use computing technology if they believe the outcome of the usage will be positive. This reflects the perception of improvement by using mobile banking apps technologies such as the speed of transactions, the convenience, ubiquitous and immediacy (Baptista and Oliveira, 2015). Performance expectancy is expected to be one of the most important factors that influence the adoption and use of mobile banking applications (Merhi et al., 2019). Therefore, the following hypothesis is postulated:

> *H1: Performance Expectancy will positively influence user behaviour towards adopting mobile banking apps*

*Effort Expectancy (EE)*

Effort expectancy is defined as *"the degree of ease associated with the use of a system"* (Venkatesh et al., 2003). In other words, the willingness of users to adopt technology is more likely when little effort is required to effectively use it. This was considered as perceived ease of use in TAM (Davis, 1989), and complexity in DOI (Rogers, 1995). Subsequently, effort expectancy has been validated in many studies that used UTAUT2 as a predictor of adoption and use of technology (Venkatesh et al., 2012; Alalwan et al., 2015; Alalwan et al., 2017). Mobile banking consumers are more inclined to accept and technology that require less effort to use. As expected, studies have revealed that there is a positive correlation between perceived ease of use of M-Banking Apps and consumers initial willingness to use it (Venkatesh et al., 2012; Aboelmaged, & Gebba, 2013; Raza, Umer, & Shah; 2017). In fact, service ease of use has been identified as the main reason consumers adopted technology (Akturan, & Tezcan, 2012). It is assumed in this study that users are more likely to adopt mobile banking apps service if it is believed to be easy to use. Therefore, the following hypothesis is proposed:

> *H2: Effort Expectancy will positively influence user behaviour towards adopting mobile banking apps*

*Social Influence (SI)*

Social Influence is *"the degree to which an individual perceives that important others believe he or she should use the new system"* (Venkatesh et al., 2003). This is because of the influence of friends, family, colleagues at work (co-workers) and the broader social networking sites (social media) on user perception and behaviour. TRA and TPB captured this as a subjective norm (Fishbein & Ajzen, 1975; Ajzen, 1991), and DOI captured it as image (Rogers, 1995). Social influence includes families, friends, co-workers, traditional media, and the social media in general. Social influence has shown to have positive effects on behavioural intentions (Akturan, & Tezcan, 2012; Venkatesh et al., 2012). People are often influenced by others to use technology, especially in the UK where technology has become part of everyday life, it is expected that friends, family, and social media become influencers of technology usage. Hence, it is hypothesised that:

> *H3: Social Influence has a positive effect on user behaviour towards adopting mobile banking apps.*

*Facilitating Conditions (FC)*

Facilitating conditions are defined as *"the degree to which an individual believes that an organisational and technical infrastructure exist to support the use of technology"* (Venkatesh et al., 2003). TRA and TPB captured this as perceived behavioural control (Fishbein & Ajzen, 1975; Ajzen, 1991), and DOI captured it as compatibility (Rogers, 1995). The use of M-Banking Apps requires some skills such as ability to install native apps, internet use experience, ability to use mobile phones or tablets as well as knowledge on security vulnerabilities. This implies that a user who has access to a set of facilitating conditions such as demos, video tutorials, manuals on mobile banking, and support chats will develop a greater intention to use the app (Baptista and Oliveira, 2015)*.* Facilitating conditions, therefore, influences intention to adopt and use mobile apps, hence the hypothesis:

> *H4a: Facilitating Conditions will positively influence user behaviour towards adopting mobile banking apps*
>
> *H4b: Facilitating Conditions will positively influence actual use behaviour of mobile banking apps*

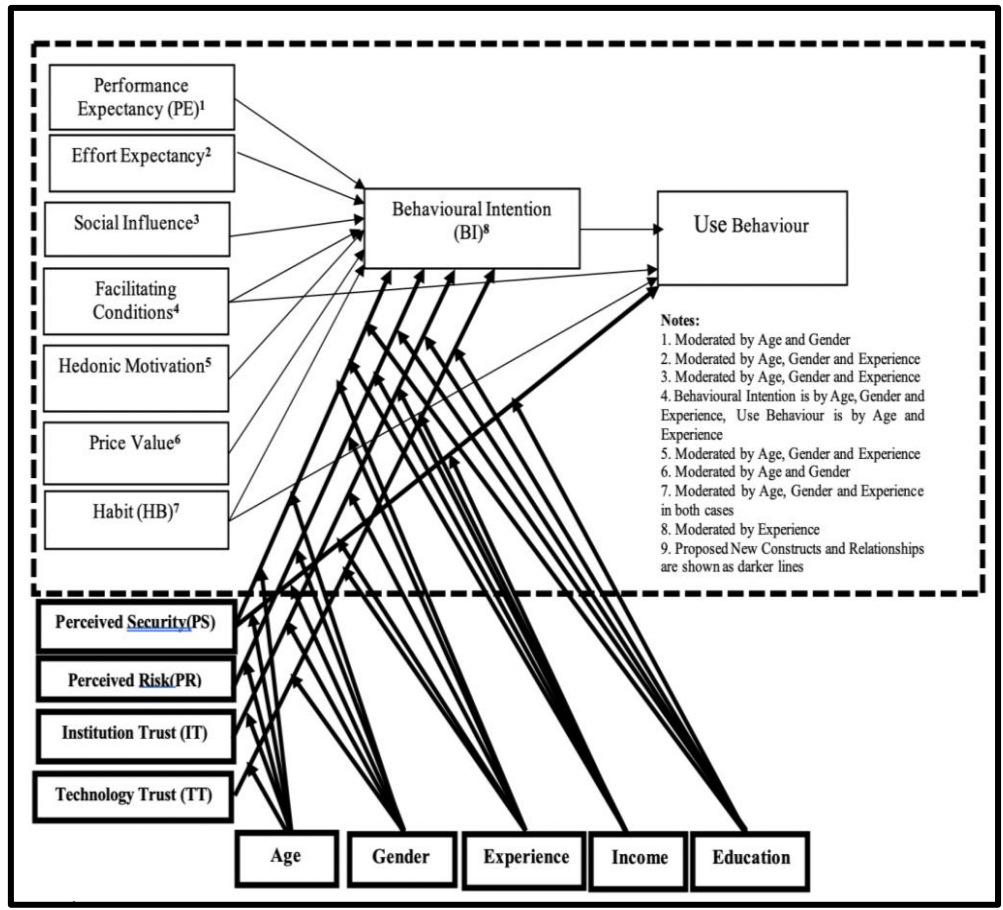

*Figure 2 Proposed Conceptual Model for the Study*

*Hedonic Motivation (HM)*

*Hedonic motivation* is the perceived enjoyment, amusement and the pleasure derived from using a technological innovation. Venkatesh *et al.* (2012) defined this as the "*fun or pleasure derived from using technology*". As the enjoyment and entertainment value of mobile banking apps increases, the greater the acceptance of use of the technology by consumers. Hedonic motivation has been found to be associated with strong mobile banking use intention (Venkatesh et al., 2012; Raza, Umer, & Shah; 2017, Aboelmaged, & Gebba, 2013). These findings and previous studies have accounted for the increase in integrating hedonic motivation for m-banking commerce and applications. Enjoyable experience and fun attached to the use of technology continually encourage mobile banking consumers to use technological-based services such as mobile banking apps. Therefore, the following hypothesis is postulated:

*H5: Hedonic Motivation positively influences behavioural intention to adopt mobile banking apps*

*Price Value (PV)*

Price value is defined as a user's cognitive trade-off between the perceived benefits of technology and the monetary cost of using it (Venkatesh et al., 2012). Price value is considered positive when the benefits associated with a technology is perceived to be greater than the cost associated with using it. According to Merhi et al. (2019), services with good price values are more likely to attract consumers. Also, pivotal relationship has been established between price of technological services and the adoption of such technology (Venkatesh et al., 2012). In the UK, cost of mobile banking service is minimal or in some cases negligible. This is considered a positive step towards the adoption of mobile banking service. Therefore, the following hypothesis is proposed:

*H6: Price Value positively influences behavioural intention to adopt mobile banking apps*

*Habit (HB)*

*Habit* deals with how people perform behaviour automatically due to learning. In effect, habit is considered as a learning behaviour with a pleasing outcome in response to automatic stimulus (Merhi et al., 2019)**.** Thus, habit will be created if consumer use technology frequently. Habit has been determined as a factor in predicting consumers behavioural intention to use m-banking technology in numerous studies (Venkatesh et al., 2012; Merhi et al., 2019; Oliveira et al., 2016; Raza et al., 2017). Consumers have been identified to use more technology when they become habituated to its use (Oliveira et al., 2016). In this study, it is anticipated that continuous use of mobile banking apps will be achieved if consumers develop positive habits towards the services. Hence, it is hypothesised that:

*H7a: Habit will have a positive influence behavioural intention to adopt mobile banking apps*

*H7b: Habit will have a positive influence on actual use behaviour of mobile banking apps*

*Perceived Security (PS)*

Data breaches that may lead to possible of data leakage, theft by cyber-criminals and hackers have contributed to security being a major concern facing mobile banking. *Perceived security* is defined as the *"degree of belief in a technology or system to transmit sensitive information without breach or leakage"* (Merhi et al., 2019)**.** Undoubtedly, security remains one of the major concerns of consumers in their use of internet banking and electronic commerce platforms (Wazid, Zeadally and Das, 2019). Cyber security vulnerabilities and its attendant challenges have left many to worry about online transactions. Adoption and use of M-Banking Apps have become a concern to consumers because of possible of data breaches and leakage, theft and damage caused by computer hackers and cybercriminals. While the UTAUT2 did not consider the aspect of security in the acceptance of technology, many other scholars and researchers have listed user security-perception as critical inhibitors of m-banking acceptance and growth of the e-banking technology (Tarhini *et al.*, 2016; Apau et al., 2019; Merhi et al., 2019). Without consumers being convinced of the security of M-Banking Apps, their trust towards the usage of such systems will be non-existent. Hence, the following hypotheses are assumed:

*H8a: Perceived Security will positively influence behavioural intention to adopt mobile banking apps with Age, Gender, Education, Income, and Experience as Moderators*

*H8b: Perceived Security will positively influence actual use behaviour of mobile banking apps with Age, Gender, Education, Income, and Experience as Moderators*

*Perceived Risk (PR)*

Perceived risk rather than perceived privacy is considered a relevant construct in this study. Contrary to other studies (Mullan, Bradley and Loane, 2017; Raza, Umer and Shah, 2017), that have argued for perceived privacy as a factor for electronic banking. This study deviates from such studies and rather considers perceived risk a critical determining factor of M-Banking app adoption. For instance, Merhi et al.(2019) hypothesised, tested, and found that perceived privacy positively influenced behavioural intention of Lebanese and British consumers to adopt mobile banking. The deviation from perceived privacy to perceived risk stems from the fact that consumers consider the risk of losing money through their mobile banking transactions as a result of security vulnerabilities (Chen, 2013; Martins, Oliveira and Popovič, 2014b; Thakur and Srivastava, 2014), than the mere disclosure of their information if that cannot result in the loss of money or cause any substantial damage. Secondly, information disclosure which perceived privacy seeks to address, is a component of security. Non-information disclosure is a parameter in the security consideration of every system (Amro and Tiantian, 2017). Additionally, the theory of perceived risk (TPR) considers many facets of risk, including privacy. In TPR, Featherman and Pavlou (2003) defined *perceived risk* as the potential for loss in the pursuit of a desired outcome in using technology or electronic service. The review has shown that perceived risk influences people's trust and security of a technological innovation (Chen, 2013; Roy *et al.*, 2017). Thus, if users perceive a technology to be risky, they will not use it. Therefore, the following hypothesis is proposed:

*H9: Perceived Risk will negatively influence behavioural intention to adopt mobile banking apps with Age, Gender, Education, Income, and Experience as Moderators*

*Institutional Trust (IT)*

Trust remains a critical antecedent that influences many electronic businesses including internet banking, mobile banking and e-commerce (Apau, Koranteng and Gyamfi, 2019). Trust of vendors and sellers in online transactions is very important. *Institutional trust* is therefore the trust between financial service providers and customers based on customers prior experience or good reputation (Merhi et al., 2019; Chiu, Chiu, & Mansumitrchai, 2016)**.** Thus, higher trust in a service provider will lower users perceived risk. Therefore, with increased trust in financial institutions and the banking sector, trust in the use of mobile banking application is expected to increase and the following hypothesis is anticipated:

*H10: Institutional Trust will positively influence behavioural intention to adopt mobile banking apps with Age, Gender, Education, Income, and Experience as Moderators*

*Technological Trust (TT)*

The desire for users to make financial transactions using mobile banking applications are influenced by the role technology plays. Technology remains an important precursor in encouraging and facilitating electronic business transactions. However, many users hesitate to transact business through electronic means due to lack of trust in the internet medium (Apau et al., 2019). *Technological trust* is the trust of users in the internet medium or the channel used for banking transactions (Apau & Koranteng, 2019)**.** Trust is also a subjective disposition to have positive assumptions for the consistent occurrence of an action (Al-Sharafi *et al*., 2018)**.** Trust of technology has been found to be a critical predictor of behavioural intention to use technology. Technology Trust is also considered an important influencer of behavioural intention due to its inverse relationship with perceived risk (Merhi et al., 2019). Therefore, higher trust in technology will lower the perceived risk and positively affects user's desire to use M-Banking Apps, hence the inclusion of trust as a new construct to be tested. Hence, the following hypothesis is proposed:

> *H11: Technological Trust will positively influence behavioural intention to adopt mobile banking apps with Age, Gender, Education, Income, and Experience as Moderators*

*Behavioural Intention (BI)*

Consistent with most technology acceptance models drawing upon psychological theories, which argue that individual behaviour is predictable and influenced by individual intentions (Yu, 2012), both the initial UTAUT and UTAUT2 support the belief and assertion that behavioural intention has substantial influence on technology use (Venkatesh et al., 2003; Venkatesh et al., 2012). Therefore, the following hypothesis is proposed:

> *H12: Behavioural Intention to adopt mobile banking apps will positively influence actual use behaviour of mobile banking apps*

Figure 3 provides detailed explanation of the hypothesised model.

## 3. Research Methodology

### 3.1 Data Collection

The survey research design is adopted for this study. In terms of time horizon, this study adopted cross-sectional rather than longitudinal. Studies on the antecedents of mobile banking adoption exist. However, the specific respondents of this study are likely to differ from previous studies, hence the use of cross-sectional. In effect, the study employed cross-sectional data collection to examine user perceptions of the security of M-Banking Apps. The use of cross-sectional research design in this study is to establish a baseline understanding of the factors that influence a user's security perception of M-Banking Apps.

Consequently, primary data collection was deemed the most feasible approach to achieve these research objectives. The proposed hypotheses for the study were tested using quantitative research approach to further understand and evaluate the applicability of the proposed conceptual model in explaining the user perceived security of M-Banking Apps. This study used questionnaire as a means of collecting quantitative primary data. The questionnaire used for data collection consisted of two parts and comprised forty-five (45) closed-ended questions. The first part comprised seven (7) closed-ended questions for the determination of demographic characteristics of study respondents using a nominal scale. These questions included age, gender, education, occupation, income, experience in using M-Banking Apps, and frequency of using M-Banking apps. The second part comprised thirty-eight (38) closed-ended questions. They included the original constructs of the UTAUT2 model. The specific questions under each construct were sourced from previous studies. Factors such as PE, SI, HM, HB, PV, and BI were measured using three items each, whereas four items were used to measure EE and FC. However, each of the extended constructs (PS, PR, IT and TT) were measured using three items. A 7-point Likert scale was adopted to measure the items used in the questionnaire. The scale ranges are 1=Strongly Disagree, 2=Disagree, 3=Somewhat disagree, 4=Neither Disagree nor Agree, 5=Somewhat Agree, 6=Agree and 7=Strongly Agree. According to Preston and Colman (2000), reliability is optimised when 7 point Likert response categories is used. The baseline UTAUT2 model also used 7 Likert scale (Venkatesh *et al.* (2012), and this will help in results comparison.

The study adopted an online survey as the means through which the questionnaire was disseminated among the target respondents. The data collection was, therefore, done distantly using Qualtrics. consequently, online questionnaire was distributed through various platforms including social media in the UK. Whilst this approach ensured that the questionnaire is widely distributed, responses were limited to UK consumers. Several approaches were employed to ensure the study findings are valid and reliable. Three participants were used to pre-test the questionnaire to ensure content validity. A pilot study was conducted using 20 participants. This was done to ensure the readability and clarity of the questionnaire items and also to verify if the collected data answered the questions under investigation by providing face validity (Brace, 2018). The pilot study was also done to ensure the conclusion and information obtained is credible for construct validity. The pilot study resulted in the removal of two questions under perceived security and the modification of two other questions.

The study adopted non-probability sampling as the technique for data collection. Specifically, the convenience sampling technique was employed. The overall consideration in selecting the sampling technique includes the size of the study population, the extent of ease of access to the study respondents as well as the homogeneity of the population. The convenience sampling also enabled data collection from potential study participants who are readily available. By using convenience sampling, any UK banking consumer is eligible to participate in the study without exception. Many researchers have adopted this sampling approach in the past for similar studies (Williams, 2018; Lim *et al.*, 2019; Merhi *et al.*, 2019).

### 3.2. Data Analysis

This study adopted Covariance-Based Structural Equation Modelling (CB-SEM) as the statistical technique for the data analysis. SEM is a statistical technique that allows simultaneous testing and estimation of a hypothesised relationships in conceptual framework (Gefen, Straub, & Boudreau, 2000). It helps to determine multiple correlations between endogenous (dependent) and exogenous (independent) variables. SEM has been identified as one of the most suitable techniques for elaborating concepts and theories without the need to resort to multiple statistical methods (Tabachnick, Fidell, & Ullman, 2007). The decision to adopt SEM for this study was based on the technique's appropriateness and applicability in situations where there is the occurrence of transition between exogenous (dependent) and endogenous (independent) variables as in the case of behavioural intentions (Wang & Wang, 2019).

Covariance-Based Structural Equation Modelling was applied using the Analysis of Moment Structures (AMOS version 26). The CB-SEM was adopted for the study due to its aptness and appropriateness for testing and confirming existing theory (Wang & Wang, 2019). The demographic characteristics of the study were analysed using descriptive statistics (Mean, standard deviation, frequency counts, Skewness, Kurtosis and Alpha value). The conceptual model was analysed using the measurement model and structural model approaches.

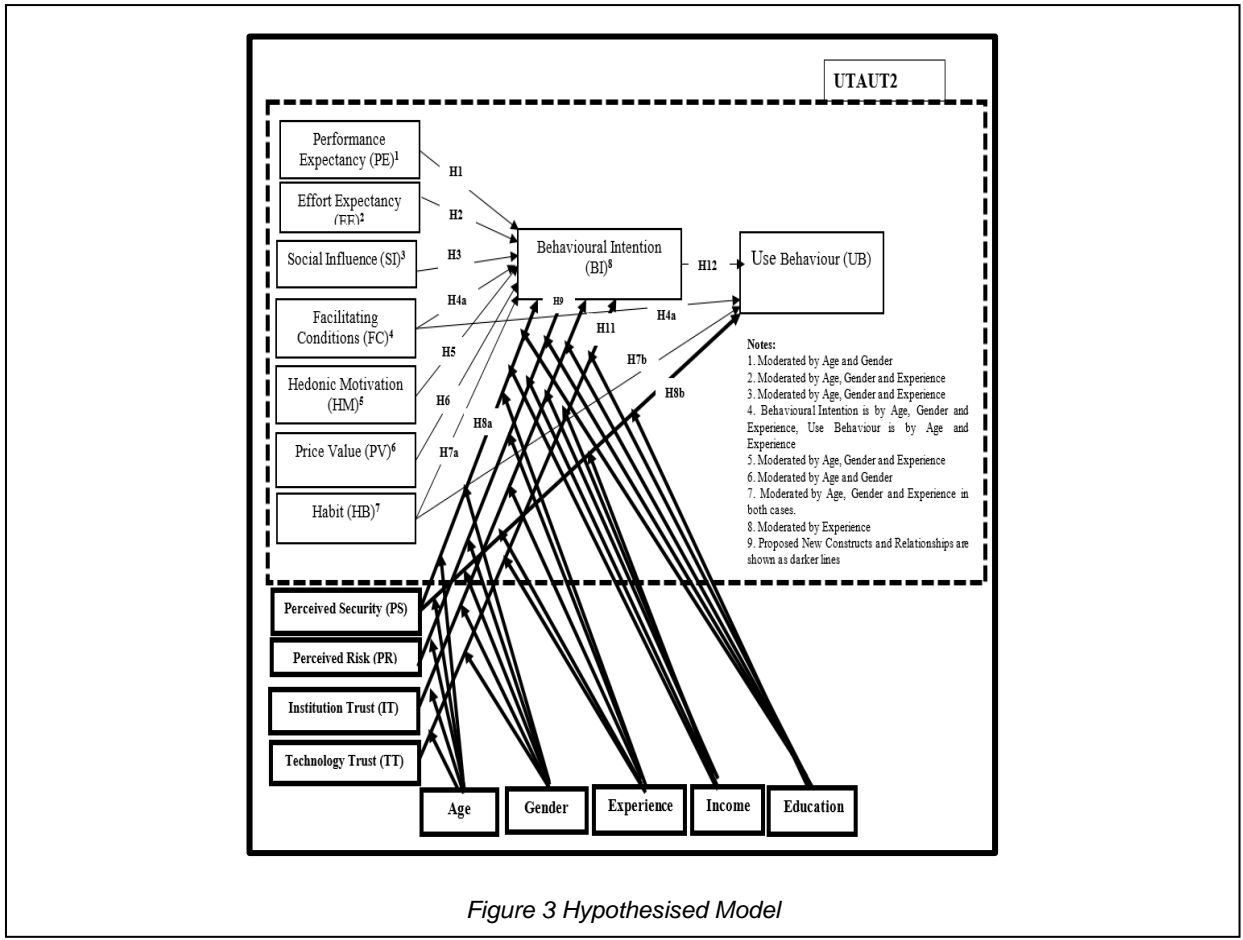

Figure 3 Hypothesised Model

## 4. Results

### 4.1. Descriptive Analysis

Data were screened for duplicate responses and missing data. There were no missing data and duplicate responses. Further treatment was therefore not required. Three (3) respondents indicated they have never used M-Banking Apps. These were removed, as the study aims to measure the experiences of M-Banking users. Consequently, 315 valid responses were retained for the final data analysis.

The initial proposed model consisted of 12 constructs and 38 indicators. These constructs are PE, EE, SI, FC, HM, HB, PV, PS, PR, IT, TT, and BI. All constructs consisted of three indicators each, except EE and FC that were measured using four indicators. The mean response for all the construct indicators was above 4 as shown in **Table 1**. The mean value ranges from 4.23 (HB3) to 6.39 (FC2). An indication that the majority of participants expressed generally positive responses to the factors in this study. A test of normality was conducted to examine whether the data is normally distributed and accurate for data analysis. The absolute kurtosis value ranges from 0.03 (HM1) to 10.169 (PE1) and absolute skewness value ranges from 0.193 (PR2) to 3.291 (PE1). The maximum kurtosis value was 10.169≈10 and the maximum skewness value was 3.291≈3. When utilising SEM, the rule of thumb suggests absolute kurtosis value of |±**10**| and skewness value of |±**3**| are considered appropriate (Kline, 2005; Brown, 2015). This means that all 36 data items in the dataset met the criteria normal distribution. Kolmogorov-Smirnova (K-S) and Shapiro-Wilk (S-W) tests were conducted to examine the sample distribution for possible response bias. Both tests were statistically significant for all sample indicating that non-response bias was not present (Ryans, 1974) (see **Table 1**).

As shown in **Table 2**, a close distribution of gender was observed. Also, the age distribution indicates that the respondents were predominantly young, aged 16-44 years (more than 90%). This possibly explains the greater level of levels of M-Banking experience recorded in this study. More than 85% of the respondents reported M-Banking usage experience of two (2) or more years. In order to better understand the use of technology in a population-based study, the participation of the older generation is valuable. However, a majority youthful participation, who often have an affinity towards technology facilitates understanding of M-banking use intentions (Merhi, Hone and Tarhini, 2019). Majority (61.91%) indicated master's degree education qualification.

**4.2. Analysis of Measurement Model**

The two step approach advocated by Anderson and Gerbing (1988) was employed to investigate the relationship in the proposed conceptual model. Firstly, confirmatory factor analysis (CFA) was employed to test the model fitness. The study employed Coltman *et al.*(2008)'s recommendations for analysing validity and reliability of the measurement model. The second step involved structural model analysis to test the proposed relationship between the exogenous variables (PE, EE, SI, FC, HM, HB, PV, PS, PR, IT, TT) and endogenous variable (BI) on one part, and exogenous variables (BI, FC, HB, PS) and endogenous variable (UB) on another part.

By employing the maximum likelihood method, the model's parameters were estimated using covariance metrices as recommended by Hair, Black, and Babin (2010). The indices examined in respect of this recommendation include *df, GFI, AGFI, CFI, RMSR, RMSEA, NFI and PNFI (Please refer to table 3 for the meaning of these statistical abbreviations)*. The sample achieved a good measurement model fit as indicated in **Table 3**.

Having achieved a good fitness model (see **Table 3**) validity and reliability of the constructs in the proposed model were evaluated. The model was assessed for composite reliability (CR), convergence validity using average variance extracted (AVE), internal consistency using Cronbach's Alpha, indicator reliability using item loadings, Kaiser Meyer Olkin (KMO), and Bartlett's test of sphericity as shown in **Table 4**. CR is employed to examine construct reliability. The recommended value for CR is 0.7 to achieve good reliability (Hair et al., 2010), and as indicated in **Table 4**, all constructs achieved CR values greater 0.7. All item loadings were above the 0.7 threshold (Thompson, Barclay, & Higgins, 1995), confirming a good indicator reliability of the instruments. The study achieved Cronbach's Alpha values greater than 0.7 as recommended by Straub (1989), suggesting constructs' validity. Convergent validity confirms the reflective nature of each construct by its own indicators, in ensuring that multiple-item factors are unidimensional (Gefen, Straub, & Boudreau, 2000) and that unreliable indicators are eliminated (Bollen, 1989). To ensure convergent validity, average variance extracted (AVE) should be above 0.5 as required by (Wixom & Watson, 2001) and

*Table 1 Results Showing Normality Test of Items*

| Items | Min | Max | Mean | SD | Skewness | Kurtosis | K-S | Df | Sig. | S-W | Df | Sig. |
|---|---|---|---|---|---|---|---|---|---|---|---|---|
| PE1 | 1 | 7 | 6.38 | 1.205 | -3.291 | 10.166 | 0.309 | 315 | 0.00 | 0.514 | 315 | 0.00 |
| PE2 | 0 | 7 | 6.07 | 1.67 | -2.448 | 5.223 | 0.348 | 315 | 0.00 | 0.578 | 315 | 0.00 |
| PE3 | 0 | 7 | 5.84 | 1.542 | -1.882 | 3.335 | 0.31 | 315 | 0.00 | 0.725 | 315 | 0.00 |
| EE1 | 1 | 7 | 6.07 | 1.036 | -2.271 | 8.77 | 0.256 | 315 | 0.00 | 0.718 | 315 | 0.00 |
| EE2 | 1 | 7 | 6.01 | 1.085 | -1.862 | 5.091 | 0.275 | 315 | 0.00 | 0.761 | 315 | 0.00 |
| EE3 | 1 | 7 | 6.04 | 1.082 | -1.608 | 3.906 | 0.242 | 315 | 0.00 | 0.784 | 315 | 0.00 |
| EE4 | 1 | 7 | 6.04 | 1.286 | -2.24 | 5.705 | 0.301 | 315 | 0.00 | 0.69 | 315 | 0.00 |
| SI1 | 1 | 7 | 5.02 | 1.607 | -0.681 | -0.22 | 0.195 | 315 | 0.00 | 0.894 | 315 | 0.00 |
| SI2 | 1 | 7 | 4.77 | 1.586 | -0.419 | -0.625 | 0.197 | 315 | 0.00 | 0.902 | 315 | 0.00 |
| SI3 | 1 | 7 | 4.67 | 1.614 | -0.39 | -0.638 | 0.177 | 315 | 0.00 | 0.916 | 315 | 0.00 |
| FC1 | 2 | 7 | 6.22 | 1.007 | -1.738 | 3.445 | 0.263 | 315 | 0.00 | 0.733 | 315 | 0.00 |
| FC2 | 1 | 7 | 6.39 | 0.89 | -3.302 | 10.169 | 0.273 | 315 | 0.00 | 0.583 | 315 | 0.00 |
| FC3 | 2 | 7 | 6.05 | 0.991 | -1.672 | 3.584 | 0.316 | 315 | 0.00 | 0.753 | 315 | 0.00 |
| FC4 | 1 | 7 | 5.47 | 1.474 | -1.255 | 1.06 | 0.306 | 315 | 0.00 | 0.821 | 315 | 0.00 |
| HM1 | 1 | 7 | 5.02 | 1.532 | -0.695 | 0.03 | 0.178 | 315 | 0.00 | 0.902 | 315 | 0.00 |
| HM2 | 1 | 7 | 5.18 | 1.495 | -0.951 | 0.361 | 0.236 | 315 | 0.00 | 0.868 | 315 | 0.00 |
| HM3 | 1 | 7 | 4.30 | 1.565 | -0.465 | -0.384 | 0.188 | 315 | 0.00 | 0.927 | 315 | 0.00 |
| HB1 | 1 | 7 | 5.56 | 1.398 | -1.438 | 1.794 | 0.304 | 315 | 0.00 | 0.804 | 315 | 0.00 |
| HB2 | 1 | 7 | 4.38 | 1.783 | 0.279 | -1.132 | 0.216 | 315 | 0.00 | 0.909 | 315 | 0.00 |
| HB3 | 1 | 7 | 4.23 | 1.89 | -0.407 | -1.042 | 0.179 | 315 | 0.00 | 0.901 | 315 | 0.00 |
| PV1 | 1 | 7 | 4.98 | 1.711 | -0.749 | -0.255 | 0.23 | 315 | 0.00 | 0.883 | 315 | 0.00 |
| PV2 | 1 | 7 | 5.40 | 1.366 | -0.712 | 0.092 | 0.224 | 315 | 0.00 | 0.88 | 315 | 0.00 |
| PV3 | 1 | 7 | 5.33 | 1.469 | -0.913 | 0.304 | 0.264 | 315 | 0.00 | 0.868 | 315 | 0.00 |
| PS1 | 1 | 7 | 5.51 | 1.324 | -1.513 | 2.406 | 0.265 | 315 | 0.00 | 0.802 | 315 | 0.00 |
| PS2 | 1 | 7 | 5.28 | 1.427 | -1.342 | 1.748 | 0.229 | 315 | 0.00 | 0.832 | 315 | 0.00 |
| PS3 | 1 | 7 | 5.51 | 1.303 | -1.458 | 2.369 | 0.254 | 315 | 0.00 | 0.814 | 315 | 0.00 |
| PR1 | 1 | 7 | 4.26 | 1.638 | 0.565 | -0.724 | 0.214 | 315 | 0.00 | 0.897 | 315 | 0.00 |
| PR2 | 1 | 7 | 4.65 | 1.687 | 0.193 | -0.958 | 0.166 | 315 | 0.00 | 0.929 | 315 | 0.00 |
| PR3 | 1 | 7 | 4.88 | 1.485 | 1.131 | 0.534 | 0.259 | 315 | 0.00 | 0.833 | 315 | 0.00 |
| IT1 | 1 | 7 | 5.25 | 1.46 | -1.40 | 1.696 | 0.248 | 315 | 0.00 | 0.814 | 315 | 0.00 |
| IT2 | 1 | 7 | 5.38 | 1.252 | -1.541 | 2.654 | 0.257 | 315 | 0.00 | 0.795 | 315 | 0.00 |
| IT3 | 1 | 7 | 5.60 | 1.088 | -1.538 | 3.558 | 0.284 | 315 | 0.00 | 0.804 | 315 | 0.00 |
| TT1 | 1 | 7 | 5.37 | 1.264 | -1.524 | 2.656 | 0.258 | 315 | 0.00 | 0.804 | 315 | 0.00 |
| TT2 | 1 | 7 | 4.53 | 1.663 | -0.409 | -1.017 | 0.193 | 315 | 0.00 | 0.905 | 315 | 0.00 |
| TT3 | 1 | 7 | 5.16 | 1.409 | -1.007 | 0.693 | 0.224 | 315 | 0.00 | 0.875 | 315 | 0.00 |
| BI1 | 4 | 7 | 6.37 | 0.68 | -0.972 | 1.178 | 0.285 | 315 | 0.00 | 0.748 | 315 | 0.00 |
| BI2 | 1 | 7 | 5.85 | 1.265 | -1.448 | 2.157 | 0.281 | 315 | 0.00 | 0.802 | 315 | 0.00 |
| BI3 | 1 | 7 | 6.02 | 1.235 | -2.052 | 4.678 | 0.304 | 315 | 0.00 | 0.711 | 315 | 0.00 |

Notes: K-S: Kolmogorov-Smirnova; S-W: Shapiro-Wilk.

the AVE should also be smaller than the CR. AVEs for the constructs were above 0.5 and smaller than CR, an indication that adequate reliability and convergent validity were achieved.

As indicated in **Table 5**, discriminant validity test was analysed based on Fornell and Lacker as well as cross-loadings criteria. Discriminant validity provides assessment for the extent to which measures different concepts are statistically significant (Gefen et al., 2000). Square roots of AVEs were computed and compared with their factor correlations values. The bold-faced values as shown in the diagonal of **Table 5**, verifying the conditions of being greater than correlations values of constructs and AVEs as recommended (Fornell & Larcker, 1981), suggesting discriminant validity. The rule of thumb further suggests that a cross factor loading of not more than 0.7 is required (Fornell & Larcker, 1981). The study assessed the possibility of multicollinearity among the exogenous constructs using bivariate correlation. The rule of thumb suggests that correlation value greater than 0.7 suggests the presence of collinearity (Tomaschek, Hendrix, & Baayen, 2018). A further test of multicollinearity was analysed by computing the variance inflation factor (VIF). VIF values less than 3 is required to avoid the

*Table 2 Demographic Characteristics of the Respondents*

| Demographic | Value | Frequency | Percentage | Cum. |
|---|---|---|---|---|
| Gender | Female | 144 | 45.71 | 45.71 |
| | Male | 171 | 54.29 | 100 |
| Age (years) | 16-24 | 45 | 14.29 | 14.29 |
| | 25-44 | 243 | 77.14 | 91.43 |
| | 45-64 | 6 | 1.9 | 93.33 |
| | 65-74 | 12 | 3.81 | 97.14 |
| | 75 and above | 9 | 2.86 | 100 |
| Education | GCSE (Level 1-2) | 6 | 1.9 | 1.9 |
| | Bachelors (Level 6) | 87 | 27.62 | 29.52 |
| | Masters (Level 7) | 195 | 61.91 | 91.43 |
| | PhD (Level 8) | 24 | 7.62 | 99.05 |
| | Others | 3 | 0.95 | 100 |
| Occupation | Academic/Teacher | 12 | 3.81 | 3.81 |
| | Clerical/Administrative | 15 | 4.76 | 8.57 |
| | Computer Technician/Engineering | 33 | 10.48 | 19.05 |
| | Executive/Manager | 27 | 8.57 | 27.62 |
| | Retired | 15 | 4.76 | 32.38 |
| | Self-employed/own company | 15 | 4.76 | 37.14 |
| | Service/Customer Support | 15 | 4.76 | 41.9 |
| | Student (college/university) | 183 | 58.1 | 100 |
| Income (Annual) | less than 3000 | 72 | 22.86 | 22.86 |
| | 3000- 10,000 | 66 | 20.95 | 43.81 |
| | 10,001- 15,000 | 63 | 20 | 63.81 |
| | 15,001- 20,000 | 42 | 13.33 | 77.14 |
| | 20,001- 25,000 | 18 | 5.71 | 82.85 |
| | 25,001 and above | 54 | 17.15 | 100 |
| Usage of Mobile Banking App | Multiple times a week | 195 | 61.9 | 61.9 |
| | Once a week | 75 | 23.81 | 85.71 |
| | Once a month | 45 | 14.29 | 100 |
| Experience in using mobile banking app | Up to one year | 45 | 14.29 | 14.29 |
| | 2-4 years | 144 | 45.71 | 60 |
| | More than 4 years | 126 | 40 | 100 |

Table 3 Fit indices summary for the measurement and structural model

| Fit Index | Recommended Values | Measurement Model | Structural Model |
|---|---|---|---|
| **Fit Index Degree of Freedom (df)** | N/A | 610 | N/A |
| **$X^2/df$** | < 5 | 2.275 | 2.076 |
| **Goodness-Of-Fit Index (GFI)** | > 0.90 | 0.978 | 1.000 |
| **Adjusted Goodness-Of-Fit Index (AGFI)** | > 0.80 | 0.901 | 0.997 |
| **Comparative Fit Index (CFI)** | > 0.90 | 0.911 | 1.000 |
| **Root Mean Square Residuals (RMSR)** | < 0.08 | 0.001 | 0.001 |
| **Root Mean Square Error of Approximation (RMSEA)** | < 0.08 | 0.017 | 0.000 |
| **Normed Fit Index (NFI)** | > 0.90 | 0.669 | 1.000 |
| **Parsimony Normed Fit Index (PNFI)** | > 0.60 | 0.868 | 0.718 |

Note: Recommended values are as required by Hair et al. (2016)

presence of multicollinearity (Hair et al., 2016). The maximum correlation value was 0.701 and the highest VIF was 2.793, suggesting multicollinearity was not present in the sample data.

Common method variance (CMV) was evaluated using Liang et al's (2007) recommended approach. The method factor computed was below 0.2, indicating common method bias is not a concern. *Post hoc* estimation of CMV was also assessed based on Malhotra, Kim, and Patil (2006) and Lindell and Whitney's (2001) approaches. A conservative estimate of the second-smallest positive correlation value (0.2) was deducted from all correlations. A re-run of the analysis showed no significant difference between original correlation values and the adjusted correlations estimates, suggesting CMV is not a problem in this study. Having achieved good constructs validity, convergence validity, indicator reliability and discriminant validity as per the results of the measurement model analysis, there is evidence of statistically distinct constructs to test the structural model.

*Table 4 Factor Loadings, Construct Reliability and Convergent Validity*

| Constructs and Indicators | Factor Loadings | Variance (%) | KMO | Bartlett's Test of Sphericity | P | Cronbach's Alpha (α) | CR | AVE |
|---|---|---|---|---|---|---|---|---|
| **Performance Expectancy (PE)** | | | 0.706 | 325.497 | 0.00 | 0.805 | 0.891 | 0.733 |
| PE1 | 0.855 | 72.903 | | | | | | |
| PE2 | 0.879 | 15.664 | | | | | | |
| PE3 | 0.826 | 11.433 | | | | | | |
| **Effort Expectancy (EE)** | | | 0.750 | 569.741 | 0.00 | 0.833 | 0.892 | 0.673 |
| EE1 | 0.722 | 67.357 | | | | | | |
| EE2 | 0.879 | 15.896 | | | | | | |
| EE3 | 0.876 | 11.753 | | | | | | |
| EE4 | 0.795 | 4.994 | | | | | | |
| **Social Influence (SI)** | | | 0.722 | 886.804 | 0.00 | 0.931 | 0.961 | 0.881 |
| SI1 | 0.896 | 87.959 | | | | | | |
| SI2 | 0.961 | 9.437 | | | | | | |
| SI3 | 0.955 | 2.605 | | | | | | |
| **Facilitating Conditions (FC)** | | | 0.671 | 218.841 | 0.00 | 0.759 | 0.812 | 0.524 |
| FC1 | 0.788 | 51.775 | | | | | | |
| FC2 | 0.831 | 19.976 | | | | | | |
| FC3 | 0.741 | 17.628 | | | | | | |
| FC4 | 0.704 | 10.62 | | | | | | |
| **Hedonic Motivation (HM)** | | | 0.702 | 513.774 | 0.00 | 0.868 | 0.922 | 0.792 |
| HM1 | 0.915 | 79.349 | | | | | | |
| HM2 | 0.921 | 14.4 | | | | | | |
| HM3 | 0.833 | 6.251 | | | | | | |
| **Habit (HB)** | | | 0.640 | 136.668 | 0.00 | 0.795 | 0.781 | 0.563 |
| HB1 | 0.777 | 55.69 | | | | | | |
| HB2 | 0.864 | 30.002 | | | | | | |
| HB3 | 0.835 | 14.308 | | | | | | |
| **Price Value (PV)** | | | 0.672 | 237.266 | 0.00 | 0.745 | 0.861 | 0.674 |
| PV1 | 0.762 | 67.225 | | | | | | |
| PV2 | 0.855 | 19.691 | | | | | | |
| PV3 | 0.840 | 13.084 | | | | | | |
| **Perceived Security (PS)** | | | 0.747 | 550.724 | 0.00 | 0.892 | 0.932 | 0.821 |
| PS1 | 0.896 | 82.323 | | | | | | |
| PS2 | 0.917 | 9.859 | | | | | | |
| PS3 | 0.909 | 7.818 | | | | | | |
| **Perceived Risk (PR)** | | | 0.733 | 491.538 | 0.00 | 0.875 | 0.923 | 0.802 |
| PR1 | 0.879 | 80.151 | | | | | | |
| PR2 | 0.915 | 11.598 | | | | | | |
| PR3 | 0.892 | 8.251 | | | | | | |
| **Institutional Trust (IT)** | | | 0.668 | 490.611 | 0.00 | 0.853 | 0.922 | 0.783 |
| IT1 | 0.853 | 78.454 | | | | | | |
| IT2 | 0.936 | 15.016 | | | | | | |
| IT3 | 0.866 | 6.53 | | | | | | |
| **Technological Trust (TT)** | | | 0.694 | 309.124 | 0.00 | 0.791 | 0.881 | 0.721 |
| TT1 | 0.861 | 71.699 | | | | | | |
| TT2 | 0.803 | 17.038 | | | | | | |
| TT3 | 0.876 | 11.263 | | | | | | |
| **Behavioural Intention (BI)** | | | 0.624 | 290.988 | 0.00 | 0.752 | 0.862 | 0.683 |
| BI1 | 0.789 | 67.696 | | | | | | |
| BI2 | 0.888 | 22.896 | | | | | | |
| BI3 | 0.876 | 9.408 | | | | | | |

Notes: KMO: Kaiser Meyer Olkin; AVE: Average Variance Extracted; CR: Composite Reliability

*Table 5 Discriminant Validity Test and Collinearity Assessment*

| | AVE | VIF | UB | PE | EE | SI | FC | HM | HB | PV | PS | PR | IT | TT | BI | GEN | AGE | EDU | INC | EXP |
|---|---|---|---|---|---|---|---|---|---|---|---|---|---|---|---|---|---|---|---|---|
| UB | N/A | N/A | N/A | | | | | | | | | | | | | | | | | |
| PE | 0.733 | 1.542 | 0.009 | 0.856 | | | | | | | | | | | | | | | | |
| EE | 0.673 | 2.103 | 0.032 | .547** | 0.820 | | | | | | | | | | | | | | | |
| SI | 0.881 | 1.539 | -0.05 | .252** | .353** | 0.939 | | | | | | | | | | | | | | |
| FC | 0.524 | 1.926 | 0.05 | .309** | .565** | .263** | 0.724 | | | | | | | | | | | | | |
| HM | 0.792 | 1.780 | -0.047 | .261** | .440** | .432** | .476** | 0.889 | | | | | | | | | | | | |
| HB | 0.863 | 1.723 | 0.103 | .215** | .256** | .429** | .113* | .406** | 0.750 | | | | | | | | | | | |
| PV | 0.674 | 1.252 | -.157** | -0.092 | 0.075 | .154** | .170** | 0.085 | .167** | 0.821 | | | | | | | | | | |
| PS | 0.821 | 2.793 | -0.02 | .199** | .327** | .386** | .253** | .243** | .370** | .321** | 0.901 | | | | | | | | | |
| PR | 0.802 | 1.496 | 0.061 | -0.026 | -0.088 | -.126* | -0.076 | -0.073 | 0.085 | -.163** | -.473** | 0.896 | | | | | | | | |
| IT | 0.783 | 2.249 | 0.044 | .153** | .354** | .216** | .258** | .260** | .266** | .251** | .658** | -.271** | 0.885 | | | | | | | |
| TT | 0.721 | 2.134 | 0.023 | .139* | .226** | .274** | .261** | .209** | .337** | .305** | .597** | -.259** | .701** | 0.849 | | | | | | |
| BI | 0.683 | 1.234 | -0.015 | .275** | .444** | .326** | .227** | .380** | .417** | .211** | .512** | -.135* | .499** | .416** | 0.826 | | | | | |
| GEN | N/A | 1.141 | 0.049 | 0.027 | -0.039 | -0.088 | .116* | -.111* | -.161** | -0.091 | -0.103 | -0.006 | -.216** | -0.106 | -.131* | N/A | | | | |
| AGE | N/A | 1.240 | -.223** | -0.07 | -0.011 | 0.02 | -.118* | 0.034 | -0.045 | 0.05 | 0.032 | -0.081 | -0.096 | -.210** | 0.044 | -0.08 | N/A | | | |
| EDU | N/A | 1.146 | .119* | .157** | 0.039 | -0.009 | 0.02 | -0.082 | 0.011 | -.157** | 0.003 | 0.078 | -.165** | -.173** | 0.078 | 0.066 | 0.004 | N/A | | |
| INC | N/A | 1.154 | .249** | 0.107 | 0.055 | 0.059 | 0.026 | -0.007 | 0.088 | -0.103 | 0.007 | -0.074 | -0.073 | -0.004 | 0.081 | .127* | .215** | 0.066 | N/A | |
| EXP | N/A | 1.118 | .116* | 0.066 | .154** | -0.073 | .170** | 0.079 | 0.07 | 0.031 | 0.036 | 0.042 | 0.074 | -0.026 | 0.087 | 0.037 | .146** | 0.041 | .122* | N/A |

Notes: *p < .05, **p < .01, ***p < .001; Factor Correlation Matrix with √AVE on the diagonal; AVE: Average Variance Extracted; VIF: Variance Inflation Factor; PE: Performance Expectancy; EE: Effort Expectancy; SI: Social Influence; FC: Facilitating Condition; HM: Hedonic Motivation; HB: Habit; PV: Price Value; PS: Perceived Security; PR: Perceived Risk; IT: Institutional Trust; TT: Technological Trust; BI: Behavioural Intention; UB: Use Behaviour; GEN: Gender; EDU: Education; INC: Income; EXP: Experience.

*Table 6 Significance of Path Coefficient*

| Relationship | Hypothesis | T Statistic | Coefficient | P Values | Results |
|---|---|---|---|---|---|
| PE-> BI | H1 | 0.727 | -0.024 | 0.468 | Not Supported |
| EE-> BI | H2 | 4.981 | 0.124* | 0.021 | Supported |
| SI -> BI | H3 | 0.701 | -0.094 | 0.484 | Not Supported |
| FC-> BI | H4(a) | -1.731 | 0.113** | 0.005 | Supported |
| FC-> UB | H4(b) | 1.061 | 0.288*** | 0.000 | Supported |
| HM-> BI | H5 | 2.501 | 0.132* | 0.013 | Supported |
| PV-> BI | H6 | 3.060 | 0.263*** | 0.000 | Supported |
| HB-> BI | H7(a) | 4.218 | 0.141* | 0.031 | Supported |
| HB-> UB | H7(b) | 2.243 | 0.354** | 0.006 | Supported |
| PS-> BI | H8(a) | 3.662 | 0.332*** | 0.000 | Supported |
| PS-> UB | H8(b) | -0.866 | 0.191* | 0.048 | Supported |
| PR-> BI | H9 | 1.213 | -0.071 | 0.226 | Supported |
| IT-> BI | H10 | 3.223 | 0.181** | 0.001 | Supported |
| TT-> BI | H11 | 2.423 | 0.218*** | 0.000 | Supported |
| BI-> UB | H12 | -0.848 | 0.432** | 0.007 | Supported |

## 4.3. Analysis of Structural Model

The study took a multi-staged approach to analyse the structural model. Specifically, the regression model building adopted the stepwise model building procedure using hierarchical regression model as recommended by Pezzullo, (2013). Four models were run separately. **Model 1** considered the constructs of the baseline UTAUT2 (PE, EE, SI, FC, HM, PV, and HB) to test the direct effects on the original model only. In **Model 2,** the UTAUT2 constructs were run with the new constructs (PS, PR, IT and TT). The **Model 3** combined the baseline constructs of UTAUT2, the new constructs and the moderators (age, gender, experience, income, and education). Lastly, the interaction terms were introduced in the fourth stepwise model (**Model 4**) to determine the effects of the interactions terms in predicting the model outcome. In each of these processes, changes in the significance relationship between independent and dependent variables, as well R-Square ($R^2$) changes were observed. Due to the complexity of the proposed relationships, spilt-sample analyses and plots were conducted in addition to the beta-coefficients, to understand the results patterns.

As indicated in **Table 6,** five constructs (EE, FC, HM, HB, and PV) of the basic structure of UTAUT2 were confirmed on BI and all the three constructs (BI, FC, and HB) were confirmed in UB. The results achieved, therefore, supports the applicability and validity of UTAUT2 as

*Table 7 Structural Model Results for Higher Order Interaction Terms*

| Interaction Terms | B | S.E |
|---|---|---|
| Dependent Variable: Behavioural Intention | | |
| TT x AGE x EDU x EXP | -1.523** | 0.056 |
| PS x GEN x AGE x EDU x INC x EXP | 2.767* | 1.14 |
| IT x GEN x AGE x EDU x INC x EXP | -1.718** | 0.637 |
| | | |
| Dependent Variable: Use Behaviour | | |
| PS x GEN x AGE x EXP | 2.344*** | 0.966 |

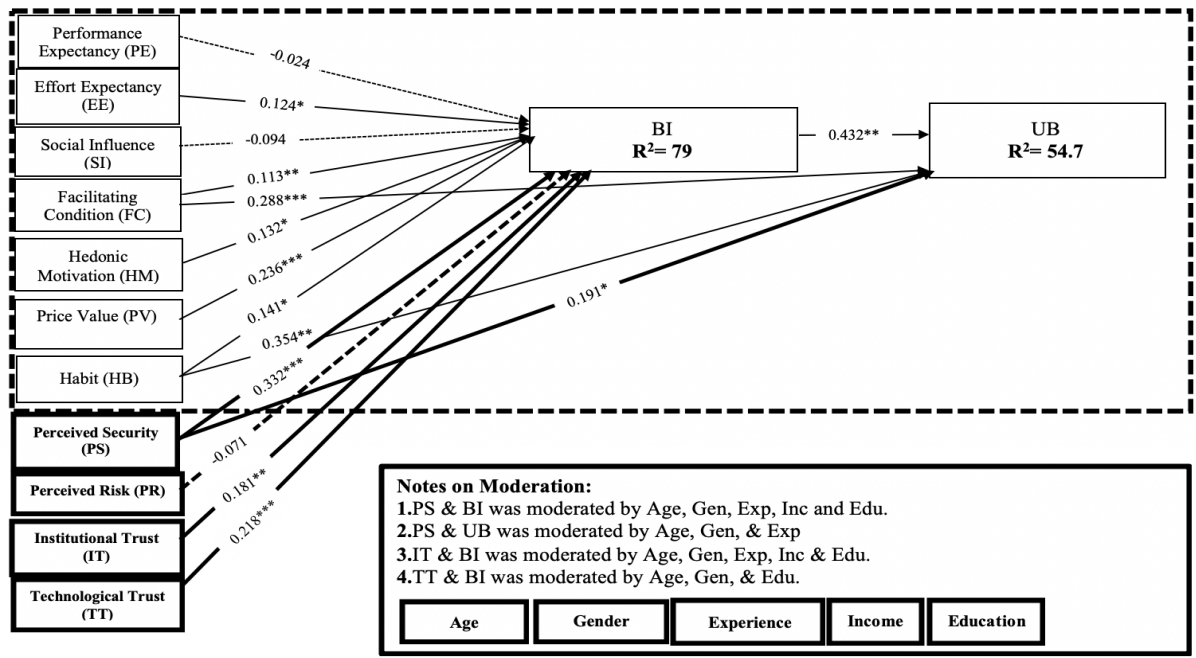

*Figure 4 Final Structural Model Results*

Notes:
1. faint broken lines show insignificant UTAUT2 constructs
2. faint straight lines show significant UTAUT2 constructs
3. Bold dark broken line shows insignificant proposed constructs

theoretical base in predicting user's behavioural intention and use behaviour of M-Banking Apps. The variance in behavioural intention and use behaviour explained by the UTAUT2 direct effects were 41.5 percent and 15.9 percent, respectively. The direct effects of the proposed model and the proposed model with interaction terms explained 43% and 79% respectively of the variance of behavioural intentions. Similarly, on the use behaviour, the variance explained 32.1% and 54.7% for direct effects only and interaction terms, respectively. **Figure 4** shows detailed results of the final structural model. The results of the higher order interactions terms are indicated in **Table 7**.

## 5. Discussion of Findings

### 5.1 Main Findings

The study presented a unique theoretical model that combined four new constructs; perceived security, perceived risk, institutional trust and technological trust with the extended unified theory of acceptance and use of technology (UTAUT2) of (Venkatesh et al., 2012), to explain user perceived security of M-Banking Apps. Theoretically, the results obtained in this study differed from that of UTAUT2 model, in that two relationships were not supported. Some of the constructs were also found to be more critical in this study than the original UTAUT2 model. The results further show that majority of the study respondents expressed positive responses to the constructs measured in this study. This supported earlier findings of Merhi *et al.* (2019), in which consumers based in England showed similar favourable responses.

As shown by the path coefficient obtained in this study, the behavioural intention to adopt M-Banking Apps was influenced by effort expectancy, facilitating condition, hedonic motivation, price value, habit, perceived security, and trust (institutional and technological). Unexpectedly, performance expectancy, social influence and perceived risk were not significant. Also, use behaviour of M-Banking Apps was influenced by behavioural intention, facilitating condition, habit, and perceived security. In terms of moderation, age, gender, experience, income, and education moderated the relationship between perceived security and behavioural intention, and the effect was stronger for younger men, particularly with high income and high education in early stages of experience with M-Banking adoption. The relationship between perceived security and use behaviour was also moderated by age, gender and experience, the effect was stronger for younger men in early stages of experience with M-Banking Apps use. The effect of institutional trust on behavioural intention was moderated by age, gender, experience, income and education, such that the effects were stronger for older women, particularly with high income and education in later stages of experience with M-Banking adoption. Meanwhile, the relationship between technological trust and behavioural intention was moderated by age, education, and experience, such that the effect was stronger for older people with high education in later stages of experience with M-Banking adoption.

The proposed model's importance as an effort to elucidate user's security perception of M-Banking Apps is clearly supported. The model does not rely only on the main relationship of the UTAUT2 but incorporates new constructs (PS, PR, IT and TT) and relationship that extends the applicability of the UTAUT2 model to M-Banking Apps context. The final model explained 79 percent and 54 percent in variance of behaviour intention and use behaviour. This is substantial, compared with baseline UTAUT2 that explained 74 percent and 52 percent of variance in intention and technology use (Venkatesh et al., 2012). The model was re-run with significant variables alone to observe the change in $R^2$. The results show that $R^2$ reduced by less than 2 percent. The majority of the significant variables in this study also achieved medium to large effect sizes. Effect sizes are by defaults 0.02, 0.15 and 0.35 indicating small, medium, and large effects respectively (Cohen, 1988). This makes the proposed model superior to the original UTAUT2 in explaining user technology behaviour.

Many scholars, including Venkatesh et al. (2012) have stressed the need and the importance to broaden the generalisability and applicability of UTAUT2 in a different context and dissimilar group to that of the original study settings. The study has effectively achieved this by supporting the applicability of the UTAUT2 theory in an M-Banking Apps use context. The literature on technology adoption as well as acceptance models and theories have, therefore, been enhanced and greatly expanded through this study. This is an indication that the proposed extensions are critical to affirming the predictive validity of UTAUT2 in examining the security perception of users in mobile banking applications use context.

Perceived security predicted behavioural intention and use behaviour of M-Banking Apps. In fact, perceived security was found to be the most significant factor and contributed the most to the variance in behavioural intention. The findings support many other earlier studies that have also found security as a major determining factor in M-banking adoption (Svilar and Zupančič, 2016; Tham *et al.*, 2017; Patel and Patel, 2018; Shareef *et al.*, 2018; Rahi and Abd. Ghani, 2018; Merhi *et al*., 2019). However, the findings contradicted those of Lim *et al.* (2019). Expectedly, younger men with a high level of education and income at the early stages of M-Banking adoption showed a stronger perception of security. This finding presupposes the mass adoption of M-Banking Apps is somewhat dependent on the overall security perception irrespective of the actual security controls that have been implemented. Security remains a big concern and barrier to M-Banking Apps due to the possibility of data breach and cyber-attacks. Therefore, reinforcement of IT infrastructure and security control mechanism remain crucial in ensuring the highest form of security in M-Banking Apps to allay the fears of users.

Trust remains one of the most important factors in online transactions. In the context of M-banking, Merhi *et al*., (2019) earlier underscored the need to categorise trust into institutional trust and the trust of technology. Both institutional trust and technological trust were found to be significant predictors of M-Banking Apps behavioural intention. These results confirmed previous findings (Muñoz-Leiva *et al.,* 2017; Tham *et al.*, 2017; Alalwan *et al.*, 2017; Stewart & Jürjens, 2018; Namahoot and Laohavichien, 2018; Merhi *et al*., 2019). However, the findings contradict other studies which did not find trust as a significant factor for behavioural intention to adopt technology (Shareef *et al.*, 2018; Williams, 2018; Alonso-Dos-Santos *et al.* 2020). While age, gender, experience, income, and education moderated the relationship between institutional trust and behavorial intention, technological trust and behavorial intention relationship was moderated by age, education, and experience. The results demonstrate the need for adequate management of customer security concerns and protecting the security of M-Banking Apps at all times. Also, detailed user-friendly reports on security implementations and improvements such as two factor authentication could be regularly desseminated. This is because, trust has emerged as an important impactful factor for M-Banking Apps adoption and has significantly demonstrated user disposition towards M-Banking technology. Also, the delicate nature of mobile financial transaction could account for the undeniable effects of trust.

The increasing trend in cyber-attacks, data compromises, and breaches (Wiafe *et al*., 2020), were expected to affect users' risk perception of mobile financial transactions. Contrary to expectations, perceived risk was not found to be a predictor of behavioural intention to adopt M-Banking Apps. Whist this finding is consistent with previous studies (Muñoz-Leiva *et al*., 2017; Williams, 2018), it is at variance with many others that found perceived risk as an important factor for mobile banking use intention (Chen, 2013; Namahoot and Laohavichien, 2018; Alonso-Dos-Santos *et al.,* 2020). Perhaps, the introduction of General Data Protection Regulations (GDPR) in 2018 has effectively mitigated the mobile banking risk concerns of banking consumers. GDPR has brought strict controls and compliance to privacy measures and institutions that fail to comply are faced with prospects of heavy fines/sanctions.

The model validated five constructs of the UTAUT2 model on behaviour intention. These are effort expectancy, facilitating conditions, hedonic motivation, price value and habits. The influence of two others; performance expectancy and social influence were not confirmed in this study. Similarly, all three constructs (behavioural intention, facilitating condition and habit) of the UTAUT2 on use technology were validated. The effort expectancy finding is consistent with earlier research (Venkatesh et al., 2012; Alalwan *et al.*, 2017; Merhi *et al*., 2019), but contradicts the findings of other studies (Im *et al.,* 2011; Baptista & Oliveira, 2015). The effort expectancy finding could be due to the increasing trend of familiarity with technology in the UK. The familiarity provides restriction for the influence of effort expectancy to users who are not well acquainted with a particular technology system. Banking institutions and policy makers should therefore continue the design of a more user-friendly interface and develop banking applications that are adaptive to different devices and sizes.

Price value was observed to be a significant predictor of behavioural intention, confirming previous studies such as Venkatesh et al. (2012) and Merhi *et al*. (2019). The finding however, contradicts Baptista and Oliveira (2015) which found the price value not to be a predictor of behavioural intention. This finding could be attributed to the low cost or absence of cost associated with the use of mobile banking applications in the UK. Contrary to studies such as Merhi *et al*. (2019), hedonic motivation was found to be a factor that influence the adoption of M-Banking Apps. This finding is consistent with previous research (Venkatesh *et al.,* 2012; Baptista & Oliveira, 2015; Alalwan *et al.*, 2017 ). This suggests that English consumers do not view M-Banking Apps as a beneficial service only but also bring them fun and enjoyment.

Habit was considered the second most important factor influencing behavioural intention and use behaviour of M-Banking Apps in line with earlier research (Venkatesh *et al.,* 2012; Alalwan *et al.*, 2017; Merhi *et al*., 2019). This result could be attributed to the established role of technology in the UK banking sector, which has made online transactions a daily routine in many cases. Similarly, facilitating conditions was found to be a significant predictor of behavioural intention and use behaviour, confirming previous studies (Venkatesh *et al.,* 2012; Alalwan *et al.*, 2017). Banks could therefore continue to provide support services to users.

Conversely, performance expectancy and social influence emerged as non-significant predictors M-Banking Apps adoption intention. This confirms an earlier study by Merhi *et al*. (2019) which reported that British consumers do not consider performance expectancy as an important factor that influences their adoption behaviour. Clearly, the decision to use technology in itself demonstrates usefulness. Perhaps, more features and services could be introduced into the Apps to make it more beneficial. Regarding social influence, the findings were consistent with previous studies (Baptista & Oliveira, 2015; Merhi *et al*., 2019). Communication regarding financial information is always kept private, which probably explains the findings.

## 5.2. Conclusion

Mobile banking applications have gained a lot of popularity within the digital space and have significantly revolutionised the banking industry. Despite the convenience offered by M-Banking Apps, users are often distrustful of the security of the applications due to increasing trend of cyber security compromises, cyber-attacks and data breaches associated with online financial transactions. Perception of security and the associated threats have increased substantially among users and innovators of mobile financial applications. This is mainly because, perceived security has been found to be an influencing factor for mobile banking usage. Security has therefore become an important consideration in M-Banking Apps development and use. Despite significant security and privacy concerns of M-Banking Apps, there is limited research in this domain.

The results indicated that most constructs of the baseline UTAUT2 structure were validated. Effort expectancy, facilitating conditions, hedonic motivation, price value and habit were found to influence behavioural intention to adopt M-Banking Apps. Similarly, facilitating conditions, habit and behavioural intention influenced M-Banking use behaviour. Unexpectedly, performance expectancy and social influence were not confirmed in this study. In terms of the extended constructs, perceived security, institutional trust and technology trust were confirmed as factors that affect user's intention to adopt and use M-Banking Apps. Perceived risk was however, not confirmed as a predictor.

The current study further revealed that in the context of mobile banking applications, the effects of security, institution and technology trust are complex. First, the impact of perceived security on behavioural intention is moderated by gender, age, experience, income, and education. Second, the effect of perceived security on use behaviour is moderated by age, gender, and experience. Third, the impact of institutional trust on behavioural intention is moderated by gender, age, experience, income, and education. Finally, the effect of technology trust on behavioural intention is moderated by age, education, and experience. Overall, the study confirmed the important role of security in influencing M-Banking Apps use, as perceived security was found to be the most important factor that predict user behavioural intention and use behaviour of M-Banking Apps.

### 5.3. Main Contribution and Advancement of Research

The major theoretical contribution of this study is in modifying UTAUT2 for mobile banking applications adoption and use context. By so doing, the generalisability of the UTAUT2 is extended from technology acceptance and use context to a more specific for M-Banking Apps. The UTAUT2 has been validated and applied in many technological settings, where performance expectancy has been the main driver of technology acceptance and use. In the specific case of M-Banking Apps, other important drivers come to play. Factors such as security, privacy, risk, and trust are important drivers in online financial transactions. Security is a critical determinant of behavioural intention of M-Banking Apps and found to be a more important driver in mobile banking use.

The study contributed to addressing a noteworthy gap in extant M-Banking Apps literature, particularly on the extension of UTAUT2 with institutional trust and technological trust. Trust remains a critical variable in online financial transactions and the addition remarkably improved the explanatory power of UTAUT2 in achieving a good model fit. The study further delineated how individual characteristics and differences such as gender, age, experience, income, and education jointly moderate the effect of security on behavioural intention and use behaviour as well as trust on behavioural intention. Interestingly, the results from this study are that the effect of security on behavioural intention are stronger in younger men with high income and education who are less experience in M-Banking Apps use, while the effect of trust was more important for older women (aged 45 years and above) with high income and education who are much experienced in M-Banking Apps use.

In summary, the proposed model incorporates not only the main relationship from UTAUT2, but also new constructs and relationships that extend the applicability of the UTAUT2 in the mobile banking application context. Empirical support has been provided for the applicability of UTAUT2 in a different cultural and industry settings in a way that offers a better explanation for the adoption and use of technology.

### 5.4. Managerial and Practical Implications

Mobile banking applications have revolutionised the banking industry. The findings of this study have several managerial and practical contributions. The results presented in this study provide relevant practical information to stakeholders and businesses in the UK financial sector as well as countries of similar cultural settings on user-perceived security of M-Banking Apps. The study has demonstrated the importance of institutional trust and technological and how they influence user's behaviour in adopting mobile banking applications. The guarantee of enhanced security, advanced privacy mechanisms and trust should be considered paramount in future strategies aimed at promoting mobile banking adoption and use of technology.

It is also recommended for mobile apps developers, software engineers, and computer programmers to integrate major security features into the development of banking apps to curtail data breaches and cyber security vulnerabilities. Institutional trust can be improved by improving brand image through celebrity endorsement, displaying positive reviews about institutions services among many others. Similarly, technology trust can also be improved by integrating features that portray the app to be trustworthy and credible. For example, showing logos on "established security institutions" in the form of third-party endorsements, letting users control their privacy settings or giving users a real-world feel by showing an "About Us" page which displays the manufacturers and bank's management improves credibility and trust. These measures, when adopted will contribute significantly by boosting the trust and confidence of the mobile banking system and further reduce the perceived risks associated with it. Improvement in the security and privacy of mobile banking apps will particularly benefit UK bank managers, as it will lead to increasing trend of adoption and use, thereby ultimately reducing in branch transactions. This will result in cost savings since operational cost associated with branch operations will be drastically reduced.

## 5.5. Limitations and Directions for Future Research

The data for this study was collected using non-probabilistic sampling technique. Specifically, the convenience sampling technique was employed for the data collection. This, therefore, places limitation on the generalisability of the study findings to the whole population. A more representative sampling approach, that eliminates possible sampling biases could be adopted in a future study. For instance, clustering and random sampling approach could be adopted to further help understand user's perception.

The complexity of the model proposed for this study made it difficult for a sub analysis of users' responses based on the category of banks whose mobile apps they use were not done. By categories, the researcher is referring to mobile banking apps developed by traditional banks (eg Barclays, HSBC, Santander, NatWest etc.) and those developed by purely modern digital banks (eg Monzo, Monese, Starling etc.). This is because, customers of these two categories have different banking experiences. While those in the traditional banks have access to branches to make transactions if they doubt mobile banking transactions, modern digital banks are purely online and branchless. It would therefore be interesting to explore how customers of these two categories perceive the security of the mobile banking apps. Future research should therefore consider this limitation by providing clear and a comprehensively extensive understanding of user-perceived security of mobile banking applications.

The study did not get enough respondents aged 45+ which ultimately may affect observations made, though initial observations indicate that the older generation has less trust in the security of M-Banking Apps. Again, the older generation particularly those over 55 years are likely to be less comfortable with adopting M-Banking Apps because of the generation gap, hence further research focusing primarily on those aged 45+ years is necessary to understand attitudes of this age group towards mobile banking applications.

Finally, future research could arrive at a better understanding of the user-perceived security of M-Banking Apps by integrating cultural dimensions. Hofstede cultural dimensions integration could be interesting, considering the unexpected results of this study such as the social influence and perceived risk. UK has a diverse cultural population, comprising people originally from typical collectivist and individualist cultural settings. Integrating cultural dimensions into the proposed model will therefore provide a better understanding of user-perceived security of M-Banking Apps.

**References**

Aboelmaged, M., & Gebba, T. R. (2013). Mobile banking adoption: an examination of technology acceptance model and theory of planned behavior. International Journal of Business Research and Development, 2(1).

Afshan, S. and Sharif, A. (2016) 'Acceptance of mobile banking framework in Pakistan', Telematics and Informatics, pp. 370–387.

Ahmed, M. S., Everatt, J. and Fox-Turnbull, W. (2017) 'Extracting Best Set of Factors that Affect Students Adoption of Smartphone for University Education: Empirical Evidence from UTAUT-2 Model', Journal of Management, Economics and Industrial Organization, pp. 51–65.

Ajzen, I. (1991) 'The theory of planned behavior', Organizational Behavior and Human Decision Processes, 50(2), pp. 179–211.

Akturan, U., & Tezcan, N. (2012). Mobile banking adoption of the youth market: perceptions and intentions, Market. Marketing Intelligence & Planning. 30 (4) (2012) 444–459.

Alalwan, A. A., Dwivedi, Y. K. and Rana, N. P. (2017) 'Factors influencing adoption of mobile banking by Jordanian bank customers: Extending UTAUT2 with trust', International Journal of Information Management, 37(3), pp. 99–110.

Alalwan, A.A., Dwivedi, Y.K., Rana, N.P. , Lal, B. , & Williams, M. D. (2015) 'Consumer adoption of Internet banking in Jordan: examining the role of hedonic motivation, habit, self-efficacy and trust', Journal of Financial Services Marketing, 20(2), pp. 145–157.

Alonso-Dos-Santos, M., Soto-Fuentes, Y. and Valderrama-Palma, V. A. (2020) 'Determinants of Mobile Banking Users' Loyalty', Journal of Promotion Management,0(0), pp. 1–19.

Al-Sharafi, A., Arshah, R.A., Alajmi, Q., Herzallah, A.T. and Qasem, Y. A. (2018) 'The Influence of Perceived Trust on Understanding Banks' Customers behaviour to Accept Internet Banking Services.', Indian Journal of Science and Technology,11(20), pp.1–9.

Ameen, N., Willis, R. and Hussain Shah, M. (2018) 'An examination of the gender gap in smartphone adoption and use in Arab countries: A cross-national study', Computers in Human Behaviour, 89, pp. 148–162.

Amro, A. and Tiantian, D. (2017) Examining young users' security perceptions of mobile banking: A qualitative study on users' insights about mobile banking. Umea University. Availavle at https://www.diva-portal.org/smash/record.jsf?pid=diva2%3A1156302&dswid=4465(accessed: 20/08/2020)

Anderson, J. C., & Gerbing, D. W. (1988) 'Structural equation modeling in practice: A review and recommended two-step approach', Psychological bulletin, 103(3), p. 411.

Apau, R and Koranteng, F. N. (2019) 'Impact of Cybercrime and Trust on the Use of E-Commerce Technologies: An Application of the Theory of Planned Behavior', International Journal of Cyber Criminology, 13(2), pp. 228–254.

Apau, R., Koranteng, F. N. and Gyamfi, S. A. (2019) 'Cyber-Crime and its Effects on E-Commerce Technologies', Journal of Information, 5(1), pp. 39–59.

Arenas-Gaitán, J., Peral-Peral, B. and Ramón-Jerónimo, M. A. (2015) 'Elderly and internet banking: An application of UTAUT2', Journal of Internet Banking and Commerce, 20(1), pp. 1–23.

Bandura, A. (1986) Social Foundations of Thought and Action: a Social Cognitive Theory Translated by Anonymous. Prentice-Hall.

Baptista, G., & Oliveira, T. (2015) 'Understanding mobile banking: The unified theory of acceptance and use of technology combined with cultural moderators.', Computers in Human Behaviour, 50, pp. 418-430.

Bhatiasevi, V. (2016) 'An extended UTAUT model to explain the adoption of mobile banking', Information Development, 32(4), pp. 799–814.

Bollen, K. A. (1989) 'A new incremental fit index for general structural equation models', Sociological methods & research, 17(3), pp. 303-316

Brace, I. (2018) Questionnaire Design: How to Plan, Structure and Write Survey Material for Effective Market Research. Kogan Page Publishers.

Brown, T. A. (2015) Confirmatory factor analysis for applied research. 2nd editio. New York: Guilford Publication.

Chauhan, V., Yadav, R., & Choudhary, V. (2021). Adoption of electronic banking services in India: an extension of UTAUT2 model. *Journal of Financial Services Marketing*, 1-14. https://doi.org/10.1057/s41264-021-00095-z

Chawla, D. and Joshi, H. (2018) 'The Moderating Effect of Demographic Variables on Mobile Banking Adoption: An Empirical Investigation', Global Business Review, 19(3_suppl), pp. 90–113.

Chen, C. S. (2013) 'Perceived risk, usage frequency of mobile banking services', Managing Service Quality, 23(5), pp. 410–436.

Chen, L. and Holsapple, C. W. (2013) 'E-business adoption research: State of the art', Journal of Electronic Commerce Research, 14(3), pp. 279–286.

Chiu, C. L., Chiu, J. L., & Mansumitrchai, S. (2016) 'Privacy, security, infrastructure and cost issues in internet banking in the Philippines: initial trust formation.', International Journal of Financial Services Management, 8(3), pp. 240–271.

Choudrie, J., Junior, C.O., McKenna, B. and Richter, S. (2018) 'Understanding and conceptualising the adoption, use and diffusion of mobile banking in older adults: A research agenda and conceptual framework.', Journal of Business Research, 88, pp. 449–465.

Cohen, J. (1988) Statistical power analysis for the behavioural sciences. 2nd edition. Hillsdale, New Jersey: Lawrence Erlbaum Associates.

Coltman, T. et al. (2008) 'Formative versus reflective measurement models: Two applications of formative measurement', Journal of Business Research, 61(12), pp. 1250–1262.

Davis, F. D. (1989) 'Perceived usefulness, perceived ease of use, and user acceptance of information technology', MIS Quarterly, 13(3), pp. 319–340.

Davis, F.D., Bagozzi, R.P., Warshaw, P. R. (1992) 'Extrinsic and intrinsic motivation to use Computers in the Workplace', Journal of Applied Social Psychology, 22(14), pp. 1111– 1132.

El-Masri, M. & Tarhini, A. (2017) 'Factors affecting the adoption of e-learning systems in Qatar and USA: Extending the Unified Theory of Acceptance and Use of Technology', Educational Technology Research and Development, 65(3), pp.743–763.

Europol. (2016) Malware has gone mobile stop think connect to keep cyber criminals out of your mobile device.Available via https://www.europol.europa.eu/newsroom/news/malware-has-gone-mobile- stopthinkconnect- to-keep-cybercriminals-out-of-your-mobile-device [Retrieved 2020- 05-15]

Fakhoury, R. and Aubert, B. (2017) 'The impact of initial learning experience on digital services usage diffusion: A field study of e-services in Lebanon', International Journal of Information Management, 37(4), pp. 284–296.

Featherman, M.S. & Pavlou, P. A. (2003) 'Predicting e-services adoption: a perceived risk facets perspective', International Journal of Human-Computer Studies, 59(4), pp. 451– 474.

Federal Reserve. (2016) Consumers and mobile financial services. [electronic]. Available via: https://www.federalreserve.gov/econresdata/consumers-and-mobile-financial-         services-report-201603.pdf [Retrieved April 15, 2020]

Fishbein, M. & Ajzen, I. (1975) 'Belief, Attitude, Intention and Behaviour: An Introduction to Theory and Research', Journal of business venturing, 5, pp. 177-189.

Fornell, C., & Larcker, D. F. (1981) 'Evaluating structural equation models with unobservable variables and measurement error', Journal of marketing research, 18(1), pp. 39-50.

Gefen, D., Straub, D., & Boudreau, M. C. (2000) 'Structural equation modeling and regression: Guidelines for research practice.', Communications of the association for information systems, 4(1), p. 7.

Hair, J. F., Black, W. C., & Babin, A. (2010) RE and Tatham, RL (2006), Multivariate Data Analysis. New Jersey: Pearson Education.

Hair, Jr., J.F., Sarstedt, M., Matthews, L.M. and Ringle, C. M. (2016) 'Identifying and treating unobserved heterogeneity with FIMIX-PLS: part I – method', European Business Review, 28(1), pp. 63-76.

Hanif, Y & Lallie, H.S. (2021). Security factors on the intention to use mobile banking applications in the UK older generation (55+). A mixed method study using modified UTAUT and MTAM- with perceived security, risk and trust. Technology in Society, 67, 101693.

Im, I., Hong, S., & Kang, M. S. (2011) 'An international comparison of technology adoption: Testing the UTAUT model', Information & Management, 48(1), pp. 1–8.

Indrawati, M. A. A., Pradhina, N. P., & Pillai, S. K. B. (2020) 'Customer continuance intention toward digital banking applications', in Understanding Digital Industry: Proceedings of the Conference on Managing Digital Industry, Technology and Entrepreneurship (CoMDITE 2019), July 10-11, 2019. Bandung, Indonesia: Routledge, p. 103.

Isaac, O., Abdullah, Z., Ramayah, T. and Mutahar, A. (2018) 'Factors determining user satisfaction of internet usage among public sector employees in Yemen', International Journal of Technological Learning, Innovation and Development, 10(1), pp. 37-68.

Karjaluoto, H., Glavee-Geo, R., Ramdhony, D., Shaikh, A.A. and Hurpaul, A. (2021), "Consumption values and mobile banking services: understanding the urban–rural dichotomy in a developing economy", *International Journal of Bank Marketing*, Vol. 39 No. 2, pp. 272-293. https://doi.org/10.1108/IJBM-03-2020-0129

Kline, R. B. (2005) Principles and practice of structural equation modelling. 2nd edition. New York: The Guilford Press.

Kumar, S., & Yukita, A. L. K. (2021, May). Millennials Behavioral Intention in Using Mobile Banking: Integrating Perceived Risk and Trust into TAM (A Survey in Jawa Barat). In *International Conference on Business and Engineering Management (ICBEM 2021)* (pp. 210-217). Atlantis Press.

Kwateng, K. O., Atiemo, K. A. O., & Appiah, C. (2019). Acceptance and use of mobile banking: an application of UTAUT2. Journal of Enterprise Information Management.

Laukkanen, T. (2015) 'How uncertainty avoidance affects innovation resistance in mobile banking: The moderating role of age and gender.', in 2015 48th Hawaii International Conference on System Sciences, pp. 3601-3610 IEEE.

Liang, H., Saraf, N., Hu, Q., & Xue, Y. (2007) 'Assimilation of enterprise systems: the effect of institutional pressures and the mediating role of top management', MIS quarterly, pp. 59-87.

Liébana-Cabanillas, F., Sánchez-Fernández, J., & Muñoz-Leiva, F. (2014) 'Antecedents of the adoption of the new mobile payment systems: The moderating effect of age.', Computers in Human Behaviour, 35, pp. 464-478.

Lim, S. H. et al. (2019) 'An Empirical Study of the Impacts of Perceived Security and Knowledge on Continuous Intention to Use Mobile Fintech Payment Services', International Journal of Human-Computer Interaction, 35(10), pp. 886–898.

Lindell, M. K., & Whitney, D. J. (2001) 'Accounting for common method variance in cross-sectional research designs.', Journal of applied psychology, 86(1), p. 114.

Luvanda, A., Kimani, S. and Kimwele, M. (2014) 'Lack of Awareness by End Users on Security Issues Affecting Mobile Banking: A Case Study of Kenyan Mobile Phone End Users', Journal of Information Engineering and Application, 4(5), pp. 19–29.

Majumdar, S., & Pujari, V. (2021). Exploring usage of mobile banking apps in the UAE: a categorical regression analysis. *Journal of Financial Services Marketing*, 1-13.

Majumdar, S., & Pujari, V. (2021). Exploring usage of mobile banking apps in the UAE: a categorical regression analysis. *Journal of Financial Services Marketing*, 1-13.

Malhotra, N. K., Kim, S. S., & Patil, A. (2006) 'Common method variance in IS research: A comparison of alternative approaches and a reanalysis of past research.', Management science, 52(12), pp. 1865-1883.

Martins, C., Oliveira, T. and Popovič, A. (2014) 'Understanding the internet banking adoption: A unified theory of acceptance and use of technology and perceived risk application', International Journal of Information Management, 34(1), pp. 1–13.

Merhi, M., Hone, K. and Tarhini, A. (2019) 'A cross-cultural study of the intention to use mobile banking between Lebanese and British consumers: Extending UTAUT2 with security, privacy and trust', Technology in Society, 59, p. 101-151.

Merhi, M., Hone, K., Tarhini, A. and Ameen, N. (2021), "An empirical examination of the moderating role of age and gender in consumer mobile banking use: a cross-national, quantitative study", *Journal of Enterprise Information Management*, Vol. 34 No. 4, pp. 1144-1168. https://doi.org/10.1108/JEIM-03-2020-0092

Mettouris, C., Maratou, V., Vuckovic, D., Papadopoulos, G. A., & Xenos, M. (2015) 'Information Security Awareness through a Virtual World: An end-user requirements analysis', in Proc. 5th International Conference on Information Society and Technology– ICIST, pp. 273–278.

Montesdioca, G. P. Z. and Maçada, A. C. G. (2015) 'Measuring user satisfaction with information security practices', Computers and Security, 48, pp. 267–280.

Mullan, J., Bradley, L. and Loane, S. (2017) 'Bank adoption of mobile banking: stakeholder perspective', International Journal of Bank Marketing, 35(7), pp. 1152–1172.

Muñoz-Leiva, F., Luque-Martínez, T. and Sánchez-Fernández, J. (2010) 'How to improve trust toward electronic banking', Online Information Review, 34(6), pp. 907–934.

Mutahar, A. M. et al. (2018) 'The effect of awareness and perceived risk on the technology acceptance model (TAM): mobile banking in Yemen', International Journal of Services and Standards, 12(2), pp. 180–204.

Namahoot, K. S. and Laohavichien, T. (2018) 'Assessing the intentions to use internet banking: The role of perceived risk and trust as mediating factors', International Journal of Bank Marketing, 36(2), pp. 256–276.

Natarajan, T., Balasubramanian, S.A. and Kasilingam, D. L. (2018) 'The moderating role of device type and age of users on the intention to use mobile shopping applications', Technology in Society, 53, pp. 79-90.

Negahban, A. and Chung, C. H. (2014) 'Discovering determinants of users perception of mobile device functionality fit', Computers in Human Behaviour, 35, pp. 75–84.

Obaid, T. (2021). Predicting Mobile Banking Adoption: An Integration of TAM and TPB With Trust and Perceived Risk. *Available at SSRN 3761669*.

Oliveira, T., Thomas, M., Baptista, G., Campos, F. (2016). Mobile payment: understanding the determinants of customer adoption and intention to recommend the technology. Computer in Human Behaviour, 61, 404–414.

Patel, K. J. and Patel, H. J. (2018) 'Adoption of internet banking services in Gujarat: An extension of TAM with perceived security and social influence', International Journal of Bank Marketing, 36(1), pp. 147–169.

Pentina, I., Zhang, L., Bata, H., & Chen, Y. (2016). Exploring privacy paradox in information-sensitive mobile app adoption: A cross-cultural comparison. Computers in Human Behavior, 65, 409-419.

Pezzullo, J. (2013) Biostatistics for dummies. John Wiley & Sons.

Picoto, W. N., & Pinto, I. (2021). Cultural impact on mobile banking use–A multi-method approach. *Journal of Business Research*, *124*, 620-628.

Prabhakaran, S., Vasantha, S., & Sarika, P. (2020). Effect of social influence on intention to use mobile wallet with the mediating effect of promotional benefits. *Journal of Xi'an University of Architecture & Technology*, *12*(2), 3003-3019.

Preston, C. C., & Colman, A. M. (2000) 'Optimal number of response categories in rating scales: reliability, validity, discriminating power, and respondent preferences.', Acta psychologica, 104(1), pp. 1-15.

Rabaa'i, A. A., & AlMaati, S. (2021). Exploring the determinants of users' continuance intention to use mobile banking services in Kuwait: Extending the expectation-confirmation model. *Asia Pacific Journal of Information Systems*, *31*(2), 141-184.

Rahi, S. and Abd. Ghani, M. (2018) 'The role of UTAUT, DOI, perceived technology security and game elements in internet banking adoption', World Journal of Science, Technology and Sustainable Development, 15(4), pp. 338–356.

Ramli, Y., Harwani, Y., Soelton, M., Hariani, S., Usman, F., & Rohman, F. (2021). The Implication of Trust that Influences Customers' Intention to Use Mobile Banking. *The Journal of Asian Finance, Economics, and Business*, *8*(1), 353-361.

Raza, S. A., Umer, A. and Shah, N. (2017) 'New determinants of ease of use and perceived usefulness for mobile banking adoption', International Journal of Electronic Customer Relationship Management, 11(1), pp. 44–65.

Raza, S. A., Umer, A., & Shah, N. (2017). New determinants of ease of use and perceived usefulness for mobile banking adoption. International Journal of Electronic Customer Relationship Management, 11(1), 44-65.

Rehman, Z. U., & Shaikh, F. A. (2020). Critical factors influencing the behavioral intention of consumers towards mobile banking in Malaysia. *Engineering, Technology & Applied Science Research*, *10*(1), 5265-5269.

Rodrigues, G., Sarabdeen, J. and Balasubramanian, S. (2016) 'Factors that Influence Consumer Adoption of E-government Services in the UAE: A UTAUT Model Perspective', Journal of Internet Commerce, 15(1), pp. 18–39.

Rogers, E. M. (1995) Diffusion of Innovations. fourth. New York: The Free Press.

Roy, Sanjit Kumar, Balaji, M. S., Ankit Kesharwani, and H. S. (2017) 'Predicting Internet banking adoption in India: A perceived risk perspective.', Journal of Strategic Marketing, 25(5–6), pp. 418–438.

Ryans, A. B. (1974) 'Estimating consumer preferences for a new durable brand in an established product class.', Journal of Marketing Research, 11(4), pp. 434-443.

Servidio, R. (2014) 'Exploring the effects of demographic factors, Internet usage and personality traits on Internet addiction in a sample of Italian university students', Computers in Human Behaviour, pp. 85–92.

Shareef, M. A. et al. (2018) 'Consumer adoption of mobile banking services: An empirical examination of factors according to adoption stages', Journal of Retailing and Consumer Services, 43, pp. 54–67.

Shareef, M. A., Baabdullah, A., Dutta, S., Kumar, V., & Dwivedi, Y. K. (2018). Consumer adoption of mobile banking services: An empirical examination of factors according to adoption stages. Journal of Retailing and Consumer Services, 43, 54-67.

Singh, S., & Srivastava, R. K. (2020). Understanding the intention to use mobile banking by existing online banking customers: an empirical study. *Journal of Financial Services Marketing*, *25*(3), 86-96.

Stewart, H. and Jürjens, J. (2018) 'Data security and consumer trust in FinTech innovation in Germany', Information and Computer Security, 26(1), pp. 109–128.

Straub, D. W. (1989) 'Validating instruments in MIS research.', MIS Quarterly, pp. 147-169.

Svilar, A. and Zupančič, J. (2016) 'User experience with security elements in internet and mobile banking', Organizacija, 49(4), pp. 251–260.

Tabachnick, B. G., Fidell, L. S., & Ullman, J. B. (2007) Using multivariate statistics. 5th edition. Boston, MA.: Pearson Education.

Tarhini, A., Alalwan, A. A., & Algharabat, R. S. (2019) 'Factors influencing the adoption of online shopping in Lebanon: an empirical integration of unified theory of acceptance and use of technology2 and DeLone-McLean model of IS success', International Journal of Electronic Marketing and Retailing, 10(4), pp. 368-388.

Tarhini, A., El-Masri, M., Ali, M. & Serrano, A. (2016) 'Extending the UTAUT model to understand the customers' acceptance and use of internet banking in Lebanon', Information Technology & People, 29(4), pp. 830–849.

Taylor, S. & Todd, P. A. (1995) 'Understanding information technology usage: a test of competing models', Information System Research, 6(2), pp. 144–176.

Thakur, R. (2018) 'Customer engagement and online reviews.', Journal of Retailing and Consumer Services, 41, pp. 48-59.

Thakur, R. and Srivastava, M. (2014) 'Adoption readiness, personal innovativeness, perceived risk and usage intention across customer groups for mobile payment services in India', Internet Research, 24(3), pp. 369–392.

Tham, J. et al. (2017) 'Internet and Data Security – Understanding Customer Perception on Trusting Virtual Banking Security in Malaysia', European Journal of Social Sciences Studies, 2(7), pp. 186–207.

Thompson, R., Barclay, D. W., & Higgins, C. A. (1995) 'The partial least squares approach to causal modeling: Personal computer adoption and use as an illustration', Technology studies: special issue on Research Methodology, 2(2), pp. 284-324.

Thompson, R.L., Higgins, C.A., & Howell, J. M. (1991) 'Personal computing: toward a conceptual model of utilization', MIS Quarterly, 15(1), pp. 124–143.

To, A. T., & Trinh, T. H. M. (2021). Understanding behavioral intention to use mobile wallets in vietnam: Extending the tam model with trust and enjoyment. *Cogent Business & Management*, *8*(1), 1891661.

Tomaschek, F., Hendrix, P., & Baayen, R. H. (2018) 'Strategies for addressing collinearity in multivariate linguistic data.', Journal of Phonetics, 71, pp. 249-267.

Touray, A. (2015) Sustainable Solutions for Last Mile Internet Access in Developing Countries: Critical Success Factors. Jyväskylä studies in computing, pp. 214.

Venkatesh, V. and Zhang, X. (2010) 'Unified theory of acceptance and use of technology: U.S. vs. China', Journal of Global Information Technology Management, 13(1), pp. 5–27.

Venkatesh, V., Brown, S.A., & Bala, H. (2013) 'Bridging the Qualitative-Quantitative Divide: Guidelines for Conducting Mixed Methods Research in Information Systems', MIS Quarterly, 37(1), pp. 21–54.

Venkatesh, V., Morris, M.G., Davis, G.B., & Davis, F. D. (2003) 'User acceptance of information technology: toward a unified view', MIS Quarterly, 27(3), pp. 425–478.

Venkatesh, V., Thong, J. Y. L., & Xu, X. (2012) 'Consumer Acceptance and Use of Information Technology: Extending the Unified Theory of Acceptance and Use of Technology', MIS Quarterly, 36(1), pp. 157–178.

Wang, J., & Wang, X. (2019) Structural equation modelling: Applications using Mplus. John Wiley & Sons.

Wang, Q., Sun, X., Cobb, S., Lawson, G., & Sharples, S. (2016) '3D printing system: An innovation for small-scale manufacturing in home settings? Early adopters of 3D printing systems in China.', International Journal of Production Research, 54(20), pp. 6017–6032.

Wazid, M., Zeadally, S. and Das, A. K. (2019) 'Mobile Banking: Evolution and Threats: Malware Threats and Security Solutions', IEEE Consumer Electronics Magazine. IEEE, 8(2), pp. 56–60.

Wechuli, N. A., Franklin, W. and Jotham, W. (2017) 'Cyber Security Challenges to Mobile Banking in SACCOs in Kenya', International Journal of Computer (IJC), 27(1), pp. 133–140.

Wiafe, I., Koranteng, F. N., Obeng, E. N., Assyne, N., Wiafe, A., & Gulliver, S. R. (2020) 'Artificial Intelligence for Cybersecurity: A Systematic Mapping of Literature', IEEE Access, 8, pp. 146598–146612.

Widyanto, H. A., Kusumawardani, K. A., & Septyawanda, A. (2020). Encouraging behavioral intention to use mobile payment: an extension of Utaut2. *Jurnal Muara Ilmu Ekonomi Dan Bisnis*, *4*(1), 87-97.

Williams, M. D. (2018) 'Social commerce and the mobile platform: Payment and security perceptions of potential users', Computers in Human Behaviour. In press.

Win, N. N., Aung, P. P., & Phyo, M. T. (2021). Factors Influencing Behavioral Intention to Use and Use Behavior of Mobile Banking in Myanmar Using a Model Based on Unified Acceptance Theory. *Human Behavior, Development And Society*, *22*(1), 19-30.

Wixom, B. H., & Watson, H. J. (2001) 'An empirical investigation of the factors affecting data warehousing success', MIS quarterly, pp. 17-41.

**Appendix: Survey**

| Construct | Indicators | Indicators Text | Source |
|---|---|---|---|
| **Performance Expectancy (PE)** | PE1 | I find mobile banking apps useful in my daily life | Venkatesh *et al.*, (2003), Venkatesh *et al.* (2012) |
| | PE2 | Using mobile banking apps help me accomplish tasks more quickly | |
| | PE3 | Using mobile banking apps increase my productivity | |
| **Effort Expectancy (EE)** | EE1 | Learning how to use mobile banking apps is easy for me | |
| | EE2 | My interaction with mobile banking apps is clear and understandable | |
| | EE3 | I find mobile banking apps easy to use | |
| | EE4 | It is easy for me to become skilful at using mobile banking apps | |
| **Social Influence (SI)** | SI1 | People who are important to me think I should use mobile banking apps | |
| | SI2 | People who influence my behaviour think that I should use mobile banking apps | |
| | SI3 | People whose opinions that I value prefer that I use mobile banking apps | |
| **Facilitating Conditions (FC)** | FC1 | I have the resources necessary to use mobile banking apps | |
| | FC3 | I have the knowledge necessary to use mobile banking apps | |
| | FC3 | Mobile banking apps are compatible with other technologies I use | |
| | FC4 | I can get help from others when I have difficulties using mobile banking apps | |
| **Hedonic Motivation (HM)** | HM1 | Using mobile banking apps is fun | Venkatesh *et al.* (2012) |
| | HM2 | Using mobile banking apps is enjoyable | |
| | HM3 | Using mobile banking apps is very entertaining | |
| **Habit (HB)** | HB1 | The use of mobile banking apps has become a habit for me | |
| | HB2 | I am addicted to using mobile banking apps | |
| | HB3 | I must use mobile banking apps | |
| **Price Value (PV)** | PV1 | Mobile banking apps I use are reasonably priced | |
| | PV2 | Mobile banking apps I use are good value for the money | |
| | PV3 | At the current price, the mobile banking apps I use provide good value | |
| **Perceived Security (PS)** | PS1 | I perceive mobile banking apps as secure | Chawla and Joshi (2018) |
| | PS2 | Mobile banking apps have rigorous security controls | |
| | PS3 | I believe that transactions through mobile banking apps are protected and secured | Merhi et al. (2019) |
| **Perceived Risk (PR)** | PR1 | The chances of losing money if I use mobile banking apps are high | Featherman and Pavlou (2003) |
| | PR2 | Internet hackers (criminals) might take control of my account if I use mobile banking apps | |
|  | PR3 | On the whole, considering all sorts of factors, it would be risky if I use mobile banking apps |  |
| **Institutional Trust (IT)** | IT1 | I trust the banks' privacy protection to the users | Chawla and Joshi (2018) |
|  | IT2 | I feel assured that banks have the legal and technological structures to protect my transactions | Merhi et al. (2019) |
|  | IT3 | I trust the banks' system will perform well and thus process my transaction correctly |  |
| **Technological Trust (TT)** | TT1 | I believe that mobile banking apps are trustworthy |  |
|  | TT2 | Even if not monitored, I trust mobile banking apps to do the right job |  |
|  | TT3 | I trust that information concerning my mobile transactions will not be known to others | Chawla and Joshi (2018) |
| **Behavioural Intention (BI)** | BI1 | I intend to use mobile banking apps in the future | Venkatesh et al., (2003), Venkatesh et al.(2012) |
|  | BI2 | I will always try to use mobile banking apps in my daily life |  |
|  | BI3 | I plan to continue to use mobile banking apps frequently |  |
| **Usage Behaviour (UB)** | UB | What is your actual frequency of use of mobile banking apps? (i) Have not used; (ii) once a month; (iii) once a week; (iv) multiple times a week | Im et al. (2011) |